\documentclass[usenatbib]{mnras}
\usepackage{amsmath,amssymb}
\usepackage[pdftex]{graphicx}

\newcommand{\dd}{{\rm d}}

\newcommand{\percc}{{\rm cm^{-3}}}

\newcommand{\E}[1]{\times 10^{#1}}

\newcommand{\zini}{z_{\rm ini}}
\newcommand{\zfin}{z_{\rm fin}}

\newcommand{\Mhalo}{M_{\rm halo}}

\newcommand{\Rhalo}{R_{\rm halo}}

\newcommand{\Mstar}{M_{\rm star}}

\newcommand{\Mmet}{M_{\rm met}}

\newcommand{\Zsun}{{\rm Z_{\bigodot}}}
\newcommand{\Msun}{{\rm M_{\bigodot}}}

\newcommand{\Zcr}{Z_{{\rm cr}}}
\newcommand{\Zth}{Z_{{\rm th}}}

\newcommand{\Tvir}{T_{\rm vir}}

\newcommand{\Mgas}{M_{\rm gas}}

\newcommand{\SRo}{\tt noIntRad}
\newcommand{\SRi}{\tt IntRad}

\newcommand{\Jcr}{J_{\rm cr}}

\newcommand{\nDCBH}{n_{\rm DCBH}}
\newcommand{\nACH}{n_{\rm ACH}}

\newcommand{\barnDCBH}{\bar n_{\rm DCBH}}
\newcommand{\barNdcbh}{\bar N_{\rm DCBH}}

\newcommand{\tth}{t_{\rm th}}
\newcommand{\fduty}{f_{\rm duty}}

\newcommand{\Mpc}{{\rm Mpc}}

\defcitealias{Chiaki15}{C15}
\defcitealias{Chon16}{C16}
\defcitealias{Chiaki17}{C17}
\defcitealias{Chiaki18}{C18}
\defcitealias{Chiaki19}{Paper I}
\defcitealias{Smith15}{S15}
\defcitealias{Hirano14}{H14}
\defcitealias{Ishigaki14}{I14}
\defcitealias{Tominaga14}{T14}
\defcitealias{Marassi14}{M14}
\defcitealias{Marassi15}{M15}
\defcitealias{Trinca22}{T22}

\usepackage{color}
\definecolor{rev}{rgb}{0.8,0.0,0.0}

\title[DCBH formation]
      {Direct-collapse black hole formation induced by internal radiation of host halos}

\author[G. Chiaki et al.]
{Gen Chiaki$^{1,2}$\thanks{E-mail: gen.chiaki@nao.ac.jp}, 
Sunmyon Chon$^{1}$,
Kazuyuki Omukai$^{1}$,
Alessandro Trinca$^{3,4,5}$,
\newauthor 
Raffaella Schneider$^{3,4,5,6}$ and
Rosa Valiante$^{4,5}$ \\
$^{1}$Astronomical Institute, Graduate School of Science, Tohoku University, Aoba, Sendai 980-8578, Japan \\
$^{2}$National Astronomical Observatory of Japan, 2-21-1 Osawa, Mitaka, Tokyo 181-8588, Japan \\
$^{3}$Dipartimento di Fisica, Sapienza Universit\`{a} di Roma, Piazzale Aldo Moro 2, I-00185 Roma, Italy \\
$^{4}$INAF/Osservatorio Astronomico di Roma, Via di Frascati 33, I-00040 Monte Porzio Catone, Italy \\
$^{5}$INFN, Sezione Roma 1, Dipartimento di Fisica, Sapienza Universit\`{a} di Roma, Piazzale Aldo Moro 2, I-00185 Roma, Italy \\
$^{6}$Sapienza School for Advanced Studies, Viale Regina Elena 291, I-00161 Roma, Italy}

\begin{document}

\date{}

\pagerange{\pageref{firstpage}--\pageref{lastpage}} \pubyear{2023}

\maketitle

\label{firstpage}

\begin{abstract}
We estimate the fraction of halos that host supermassive black holes (SMBHs) forming through the
direct collapse (DC) scenario by using cosmological $N$-body simulations combined with a semi-analytic model for galaxy evolution. 
While in most of earlier studies the occurrence of the DC is limited only in chemically pristine halos, we here suppose that the DC can occur also in halos 
with metallicity below a threshold value $\Zth= 0$--$10^{-3}~\Zsun$, considering the super-competitive accretion pathway for DC black hole (DCBH) formation.
In addition, we consider for the first time the effect of Lyman-Werner (LW) radiation from stars within host halos, i.e., internal radiation. 
We find that, 
with low threshold metallicities of $\Zth \leq 10^{-4}~\Zsun$, 
the inclusion of internal radiation rather reduces the number density of DCBHs from $0.2$--$0.3$ to $0.03$--$0.06~\Mpc ^{-3}$.
This is because star formation is suppressed due to self-regulation, and the LW flux emitted by neighboring halos is reduced.
Only when $\Zth$ is as high as $10^{-3}~\Zsun$, internal radiation enhances the number density of DCBHs from $0.4$ to $1~\Mpc ^{-3}$, thereby decreasing the threshold halo mass above which at least one DCBH forms from $2\E{9}$ to $9\E{8}~\Msun$.
We also find that halos with $\Mhalo \gtrsim 10^{11}$--$10^{12}~\Msun$ can host more than one DCBH at $z=0$.
This indicates that the DC scenario alone can explain the observed number of SMBH-hosting galaxies.
\end{abstract}



\section{INTRODUCTION}

Supermassive black holes (SMBHs) in a wide range of masses $10^5-10^{10} \Msun$ are ubiquitously observed in galaxies \citep{Gehren84, Filippenko03, Miller15}.
Even those as massive as $\gtrsim 10^{9}~\Msun$ are found as bright quasars at redshifts of $z\gtrsim 6$, less than a billion years after the Big Bang \citep{Mortlock11, Wu15, Wang21}.
Their early appearance poses a puzzle in modern astronomy.

It is unlikely that such a gigantic object forms out of nowhere, researchers usually suppose that a somewhat smaller seed BH forms first and then grows into a SMBH.
Some scenarios have been proposed for the emergence of the high-redshift SMBHs \citep[][for a review]{Inayoshi20}.
Growing to SMBHs from the remnant BHs of the first-generation of stars ($\sim 100~\Msun$) requires continuous (or even mildly super-)Eddington-limited gas accretion for several decades in mass. 
This is not very likely because strong feedback from the stars and accreting BHs may blow up the ambient gas reservoir from the shallow potential well of their low-mass ($\sim 10^5$--$10^6~\Msun$) host halos \citep{Johnson07, Smith18, Pfister19}.
Another scenario invokes more massive BHs ($\sim 10^5~\Msun$) as the seeds.
As an origin of such massive seeds, the formation of supermassive stars (SMSs) with masses of $\sim 10^5-10^6~\Msun$ followed by direct collapse (DC) attracts attention of researchers in the last decade \citep{Latif13, Chon18}.
In a current popular variant of the scenario \citep{Bromm03}, a SMS is supposed to form as a result of monolithic collapse of a massive primordial-gas cloud irradiated by an intense Lyman-Werner (LW) radiation field (11.2--13.6 eV), which dissociates H$_2$ and disables its cooling \citep{Omukai01}: 
without H$_2$, the cloud cools by the H Lyman $\alpha$ emission and  
collapses quasi-isothermally at $\sim 8000$ K without fragmentation. 
To keep such a hot gas cloud gravitationally bound, the halo hosting the cloud must be sufficiently massive with virial temperature $\Tvir \gtrsim 8000$ K, or, in other words, it must be 
an atomic-cooling (AC) halo.
Other effects such as gas heating accompanied by halo mergers \citep{Wise19, Regan20} or delayed virialization due to streaming velocities \citep{Tanaka14, Hirano17} may work in concert with the strong radiation field for successful collapse without fragmentation by prohibiting the temperature in the cloud to drop significantly. 


In those earlier works, most authors assumed that SMSs form only in chemically pristine environments by speculating that fragmentation of clouds caused by dust cooling prohibits their formation in non-zero metallicity cases \citep{Omukai08}.
Recently, however, \citet{Chon20} demonstrated that
SMSs can still form in a massive, strongly irradiated cloud with different degrees of metal-enrichment with numerical simulations.
When the metallicity is above $\sim 10^{-5}~\Zsun$, the dust cooling becomes effective and indeed causes vigorous fragmentation, as expected previously.
The global gas inflow, however, efficiently feeds with mass a small number of massive stars residing at the center of the forming stellar system.  
Consequently, the central objects become supermassive through the so-called {\it super-competitive accretion} regardless of fragmentation occurring at such high densities as the dust cooling is effective ($\ga 10^{11}{\rm cm^{-3}}$) as long as the metallicity is below a threshold value $\Zth \sim 10^{-4}-10^{-3}~\Zsun$. 
For this reason, we consider here the formation of a SMS and a subsequent DCBH even in a halo with a finite degree of metal enrichment.

In addition, earlier works payed more attention to SMS/DCBH formation in significantly ($3$--$5\sigma$) overdense regions with the highest-$z$ quasars in mind \citep{Valiante16, Chon16, Li21}.
In those overcrowded regions, the close proximity of halos each other elevates the LW intensities at the halos and facilitates DCBH formation.  
SMBHs, however, exist not only in luminous quasars but also in normal galaxies ubiquitously in the local universe.
In fact, the occupation fraction of SMBHs in galaxies, i.e., the fraction of galaxies that host SMBHs, is known to approach unity for $\Mstar \gtrsim 10^9~\Msun$ \citep{Miller15} from X-ray observations or velocity dispersion measurement of galactic bulges \citep[see also][]{Trinca22, Spinoso22}.

An improvement in our study is the inclusion 
of radiation sources inside the same halo.
In most previous studies, strong LW radiation at SMS-forming sites is supposed to be provided by neighboring halos, i.e., {\it external radiation} sources.
Stars/galaxies in the same halo, however, should also contribute to the LW radiation in it, i.e., {\it internal radiation}.
In fact, numerical simulations of \citet{Dunn18} showed that, in a halo hosting multiple star-forming clumps, the LW radiation from stars in one clump can induce DCBH formation in another clump.
Several authors also have reported multiple clump formation in DCBH-host halos in their cosmological simulations \citep[e.g.,][]{Latif14, Latif22, Regan20}.
We thus include the effect of internal radiation in our semi-analytic modelling
and compare the efficiency of DCBH formation in cases with and without the internal radiation.

The paper is organized as follows:
In Section \ref{sec:methods}, we describe our numerical schemes of
cosmological $N$-body simulations and semi-analytical modelling for galaxy evolution.
In Section \ref{sec:results}, we present the results for the halo evolution and the number density of DCBHs.
We discuss limitation and caveats in our study in Section \ref{sec:discussion}.
We conclude the paper in Section \ref{sec:conclusion}.

In the simulations, we adopt the cosmological parameters 
$\Omega _{\rm m} = 0.3086$,
$\Omega _{\rm b} = 0.0483$,
$\Omega _{\rm \Lambda} = 0.6914$,
$h = 0.6777$ and
$\sigma _8 = 0.8288$ \citep{Planck14}.
Throughout this paper, we describe physical quantities in comoving coordinates,
unless otherwise specified.


\begin{figure}
\includegraphics[width=0.5\textwidth]{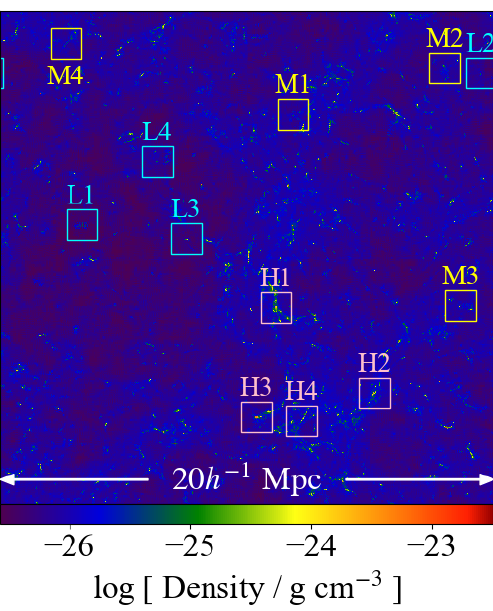}
\caption{The density projection map (density-weighted)  of dark matter at redshift $z=10$.
We plot the whole $N$-body simulation box with a comoving side length of $20 h^{-1}$ Mpc.
The red, yellow and blue rectangles indicate the high-, intermediate- and
low-density zoom-in regions, respectively, with their IDs.
Note that the region {\tt L2} is plotted on the both sides along the $x$-axis due to the periodic boundaries.}
\label{fig:proj_all_cic_all_cic_0045}
\end{figure}

\begin{table}
\begin{minipage}{\columnwidth}
\caption{Overdensities of the zoom-in regions at $z=10$}
\label{tab:zoomin}
\begin{tabular}{cccccc}
\hline
Halo & $\Mhalo$  & $\Rhalo$ & $1+\delta$ & Probability & Rarity \\
     & [$\Msun$] &  [pc]    &                    &       & [$\sigma $] \\
\hline                                                          
{\tt H1} &  $2.13\E{10}$ &         $404$ &        $3.32$ &  $4.84\E{-6}$ &        $4.42$ \\
{\tt H2} &  $2.12\E{10}$ &         $404$ &        $2.40$ &  $5.33\E{-4}$ &        $3.27$ \\
{\tt H3} &  $3.97\E{10}$ &         $498$ &        $2.32$ &  $8.34\E{-4}$ &        $3.14$ \\
{\tt H4} &  $1.82\E{10}$ &         $384$ &        $2.09$ &  $2.75\E{-3}$ &        $2.78$ \\
{\tt M1} &   $2.53\E{9}$ &         $199$ &        $1.78$ &  $1.39\E{-2}$ &        $2.20$ \\
{\tt M2} &   $8.44\E{8}$ &         $138$ &        $1.78$ &  $1.39\E{-2}$ &        $2.20$ \\
{\tt M3} &   $2.17\E{9}$ &         $189$ &        $1.78$ &  $1.39\E{-2}$ &        $2.20$ \\
{\tt M4} &   $4.82\E{8}$ &         $114$ &        $1.78$ &  $1.40\E{-2}$ &        $2.20$ \\
{\tt L1} &   $2.41\E{8}$ &        $90.8$ &        $1.07$ &       $0.346$ &       $0.395$ \\
{\tt L2} &   $3.62\E{8}$ &         $104$ &        $1.01$ &       $0.434$ &       $0.165$ \\
{\tt L3} &   $4.22\E{8}$ &         $109$ &       $0.992$ &       $0.455$ &       $0.112$ \\
{\tt L4} &   $1.21\E{8}$ &        $72.1$ &       $0.979$ &       $0.474$ &  $0.0652$ \\
\hline
\end{tabular}
\medskip \\
Note: (1) ID of zoom-in regions,
(2, 3) Mass and radius of the central halos,
(4--6) Overdensity, probability and rarity of the zoom-in regions.
\end{minipage}
\end{table}

\begin{figure}
\includegraphics[width=0.5\textwidth]{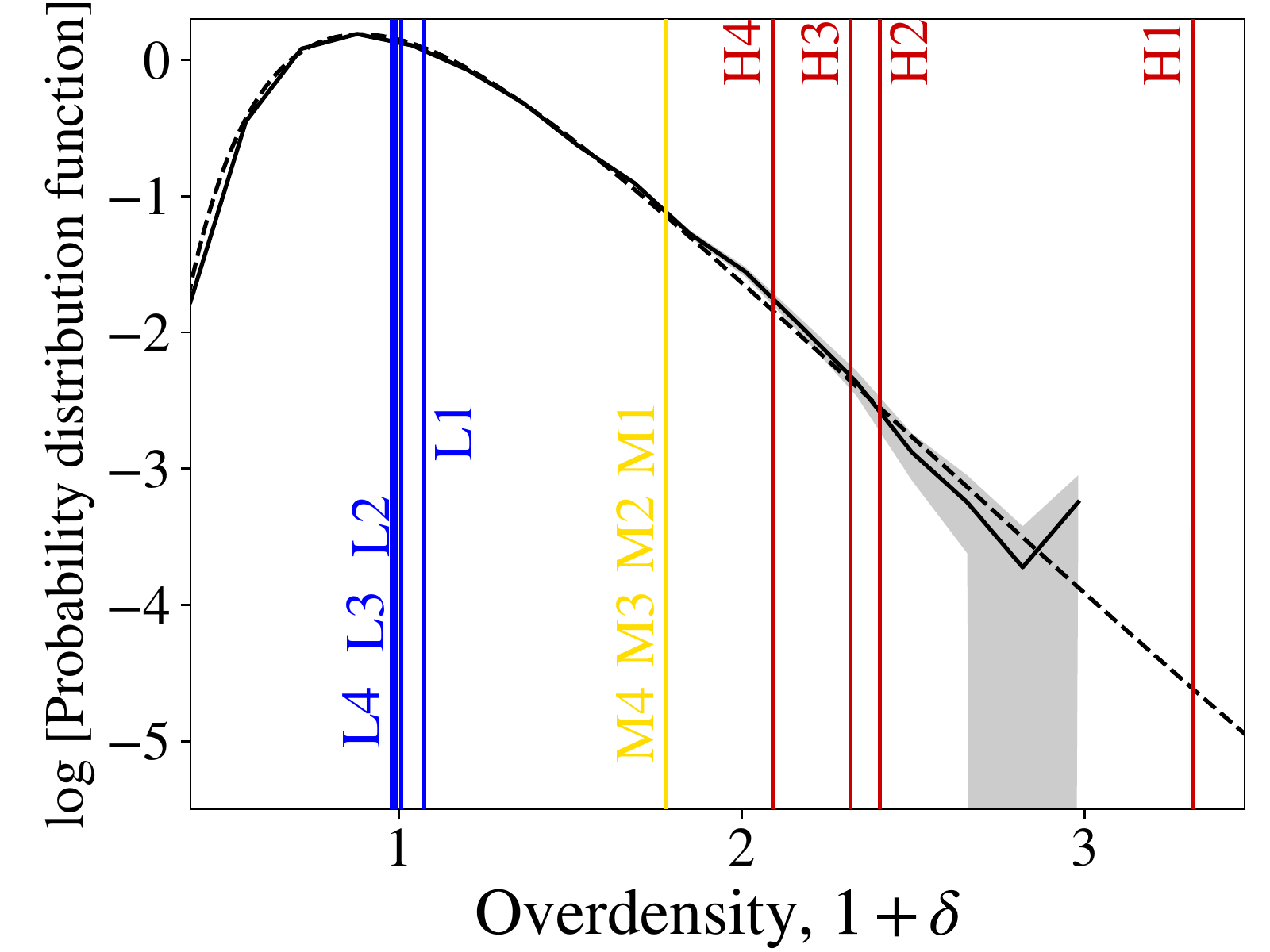}
\caption{The distribution function of matter overdensity, $1+\delta$, at the final snapshot of our simulation, $z=10$.
The black solid curve shows the result of our $N$-body simulation devided into $32^3$ regions of $1.24~\Mpc$ on each side. 
The uncertainty is indicated by the grey shaded region.
The fractional distribution is fitted with a log-normal function
(black dashed curve) with mean $\mu=-0.34$ and standard deviation $\sigma=0.28$. 
We also indicate the overdensity of the zoom-in regions with the vertical lines.
}
\label{fig:od0045}
\end{figure}

\section{Methods}
\label{sec:methods}

We carry out cosmological $N$-body simulations coupled with 
a semi-analytical model for galaxy evolution to estimate how many DCBHs are formed.
In this section, we describe our numerical approach.

\subsection{Cosmological zoom-in simulations}
\label{sec:zoomin}
We solve the dynamics of DM particles
by using the $N$-body module of 
the adaptive mesh refinement (AMR)/$N$-body simulation code {\sc enzo} 
\citep{Bryan14, BrummelSmith19}.\footnote{While {\sc enzo} also has a hydrodynamics module, we here only use the $N$-body module.}  
We initialize the simulations in a $20 h^{-1}~\Mpc$ box on each side at redshfit $\zini = 99$.
The number of DM particles is $512^3$, which corresponds to a mass resolution of $4.08\E{7}~\Msun$.
For easier comparison with the results of our previous work, \citet[][hereafter \citetalias{Chon16}]{Chon16}, we use the same initial condition produced by the initial condition generator {\sc music} \citep{Hahn11}.

Since the resolution 
of the above mentioned simulation is insufficient for resolving minihalos of $\sim 10^5$--$10^6~\Msun$, which may host Pop III star formation, 
we carry out zoom-in simulations for 12 DM halos selected 
at redshift $\zfin = 10$ in the following way:  
we first trace back the positions of Lagrangian DM particles within twice the virial radius of each halo at $\zini$ and refine DM particles in a cube with a side length of $1.24h^{-1}~\Mpc$ around the particles' center of mass with effective resolution of $8162^3$, corresponding to the minimum DM particle mass $1.25\E{3}~\Msun$.
We then re-simulate until $\zfin$.
This ensures that each minihalo is resolved with $\gtrsim 100$ particles.
To see environmental effects on DCBH formation, we take regions with various degrees of overdensity 
$1 \lesssim 1+\delta \lesssim 3$ as the zoom-in regions. 
Here, as usual, $\delta = \rho / \bar \rho - 1$, with the matter density around a halo $\rho$ and the critical density $\bar \rho$ at redshift $z$.  

The properties of our zoom-in regions are summarized in Table \ref{tab:zoomin} and their location at $\zfin$ is shown in Fig. \ref{fig:proj_all_cic_all_cic_0045}.
The 12 zoom-in regions can be classified into three sets of different overdensities, each of which consists of four regions.
The first set ({\tt H} regions) is of regions with high overdensities of $1+\delta \gtrsim 2$, tagged {\tt H1}, {\tt H2}, {\tt H3} and {\tt H4} in the order of overdensity, which are centered at the most massive halos with $\Mhalo \gtrsim 10^{10}~\Msun$ at redshift $\zfin$, locating at a crossing point of two dense filaments in large-scale structure (Fig. \ref{fig:proj_all_cic_all_cic_0045}).
The second ({\tt M} regions) and third set ({\tt L} regions) are of intermediate-density regions ({\tt M1-4}) and of low-density regions ({\tt L1-4}) with their median overdensities $1+\delta = 1.78$ and $0.98 \leq 1+\delta \leq 1.07$, respectively.
The 12 halos are selected so that the zoom-in regions do not spatially 
overlap with each other.

To compare with observations later (in Section \ref{sec:Ndcbh}), we measure the probability with which a region has an overdensity $\delta$.
We divide the entire simulation box into $32^3$ evenly-spaced cubes with a side length of $1.24 h^{-1}~\Mpc$ and calculate the overdensity of each cube.
Fig. \ref{fig:od0045} shows the probability distribution of overdensity of the sampled regions (black solid curve) with standard deviation of a Poisson distribution (grey shade).
The distribution can be fitted with a log-normal function
\begin{equation}
f(1+\delta) = \frac{1}{\sqrt{2\pi} \sigma (1+\delta)}
\exp \left\{ -\frac{[\ln (1+\delta) - \mu]^2}{ 2\sigma ^2} \right\}
\label{eq:od}
\end{equation}
with mean $\mu = -0.34$ and standard deviation $\sigma = 0.28$, which is shown by the black dashed curve.
The overdensities of the zoom-in regions
are indicated by the vertical lines. 
%
From Eq. (\ref{eq:od}), we can estimate the probability $p(1+\delta)$ for regions to have overdensity above $1+\delta$.
We also define the ``rarity'' $\nu (1+\delta)$ in units of $\sigma $ as
\begin{equation}
\nu (1+\delta) = \frac{\ln (1+\delta) - \mu}{\sigma } .
\end{equation}
The probability 
$p(1+\delta)$ and the rarity $\nu (1+\delta)$ are shown in the 5th and 6th columns of Table \ref{tab:zoomin}, respectively.
This indicates that the highest overdensity regions ({\tt H1}--{\tt 4}) are rare ($3$--$4\sigma$) while
the lowest overdensity regions ({\tt L1}--{\tt 4}) are normal ($\sim 0\sigma$).

We terminate the zoom-in simulations at redshift $\zfin = 10$.
We output snapshots at every $\Delta t_{\rm out} = 10$ Myr and
obtain 47 snapshots in each run.
We construct merger trees with {\sc consistent-trees} \citep{Behroozi13b} by identifying halos with the algorithm {\sc rockstar}, where friend-of-friend groups are identified in the six-dimensional phase space \citep{Behroozi13a}.

\subsection{Semi-analytical star formation model}

Here we describe our semi-analytic model to follow
the time evolution of the gas, star and metal masses in each halo, which is based on \citet{Salvadori07, Salvadori08} and \citet{Valiante11, Valiante16}.

\subsubsection{Pop III star formation}


\citet{Machacek01} gave 
the minimum halo mass for Pop III star formation as a function of LW intensity $J_{21}$ (in units of 
$10^{-21}$ erg s$^{-1}$ cm$^{-2}$ Hz$^{-1}$ sr$^{-1}$): 
\begin{equation}
M_{\rm th, M01} = \psi \left( 1.25\E{5} + 2.8\E{6} J_{21} ^{0.47} \right) \Msun ,
\end{equation}
where a correction factor $\psi = 4$ from \citet{OShea08}.
This does not consider the redshift dependence of the minimum halo mass in the absence of the radiation. 
We here replace the first term with the mass corresponding to a virial temperature of $\Tvir = 2000$ K \citep{Bryan98} as 
\begin{equation}
M_{\rm th, 0}(z) = 
1.0\E{6}~\Msun
\left( \frac{\Tvir}{2000~{\rm K}} \right) ^{3/2}
\left( \frac{1+z}{20} \right) ^{3/2} .
\end{equation}
Then, the minimum mass for star formation is
\begin{eqnarray}
M_{\rm th}(z, J_{21}) = M_{\rm th, 0}(z) 
+ 2.8\E{6} ~\Msun \psi  J_{21} ^{0.47} .
\label{eq:Mth}
\end{eqnarray}
We assume Pop III star formation occurs only in halos with metallicities below a critical value, for which we adopt $\Zcr = 10^{-5.5}~\Zsun$ \citep{Tsuribe06, Tsuribe08, Chiaki15}.
With metallicities above $\Zcr$, we assume that Pop II stars are formed, with a smaller characteristic stellar mass (see Section \ref{sec:PopI_II}) due to fragmentation of star-forming clouds induced by the dust cooling \citep{Omukai00, Schneider03, Omukai05}.
When the condition above is satisfied for a halo, we assume that just one Pop III star forms in it, whose mass is stochastically assigned following the Larson-type initial mass function (IMF)
\begin{eqnarray}
\Phi _{\rm PopIII} (m) &\propto & m^{-2.3} \exp \left[-\left( \frac{m_{\rm ch}}{m} \right)^{1.6} \right] ,
\end{eqnarray}
with the characteristic mass $m_{\rm ch} = 20~\Msun$ and 
the minimum and maximum masses
$(m_{\rm min},~m_{\rm max}) = (1,~300~\Msun)$, respectively.

A massive Pop III star explodes as a SN after its lifetime and releases the energy and metals promptly into its surrounding.
We take the lifetimes from \citet{Schaerer02}, 
and the explosion energy, ejecta mass and metal yield from \citet{Nomoto06}, as a function of 
the progenitor mass $m$.
Specifically, stars with $8 < m/\Msun < 40$, explode as core-collapse supernovae (CCSN) with explosion energy $1\E{51}$ erg and metal mass $\sim 1~\Msun$.
Stars with $140 < m/\Msun < 260$ explode as pair-instability supernovae (PISN) with higher explosion energy $\sim 20\E{51}$ erg and higher metal mass $\sim 100~\Msun$.
In the other mass ranges, stars directly collapse into BHs with no ejection of the material.

\subsubsection{Pop I/II star formation and chemical evolution}
\label{sec:PopI_II}

We calculate the total masses of PopI/II stars $\Mstar$, of the gas $\Mgas$ and of metals $\Mmet$ in a halo with mass $\Mhalo$, by the following equations:
\begin{eqnarray}
\frac{\dd \Mstar}{\dd t} &=& R_{\rm SF} - R_{\rm ej} , 
\label{eq:dotMstar} \\
\frac{\dd \Mgas}{\dd t} &=& -R_{\rm SF} + R_{\rm ej}
+ \frac{\dd M_{\rm inf}}{\dd t} - \frac{\dd M_{\rm ej}}{\dd t} , 
\label{eq:dotMgas} \\
\frac{\dd M_{\rm met}}{\dd t} &=&
-Z_{\rm halo} R_{\rm SF} + \frac{\dd Y}{\dd t} \nonumber \\ &&
+ Z_{\rm IGM} \frac{\dd M_{\rm inf}}{\dd t}
- Z_{\rm w} \frac{\dd M_{\rm ej}}{\dd t} 
\label{eq:dotMmet}
\end{eqnarray}
with a time interval of $\Delta t = 1$ Myr.

The first term in the right hand side of Eqs. (\ref{eq:dotMstar}) and
(\ref{eq:dotMgas}) represents the depletion rate of gas into stars, which is given by
\begin{equation}
R_{\rm SF}(t) = \epsilon _* \frac{\Mgas}{t_{\rm acc}} ,
\end{equation}
where $\epsilon _*$ is the star formation efficiency and set to $0.045$, as in the most quiescent 
star formation model of \citet{Valiante11}, 
and the gas accretion time
\begin{equation}
t_{\rm acc} = \frac{\Mhalo }{\dot M _{\rm halo}} ,
\end{equation}
is assumed to be the same as that of the host halo.
The timescale can be negative, for example, when a satellite halo is tidally disrupted by the main halo just before merger.
In such a case, we set $R_{\rm SF} = 0$. 

The second term represents the ejection rate from the star to gas due to SN explosions, which is given by
\begin{equation}
R_{\rm ej} (t) = 
\int _{0} ^{t}  \frac{\dd f_{\rm ej}}{\dd t} (t - t^{\prime}) R_{\rm SF} (t^{\prime}) \dd t^{\prime} ,
\end{equation}
where the mass fraction of gas ejected by SNe 
\begin{equation}
\frac{\dd f_{\rm ej}}{\dd t} (t) = 
\frac{m_{\rm ej} (m_t) \Phi (m_t) }{\int _{m_{\min}} ^{m_{\max}} m \Phi (m) \dd m}
\frac{\dd m_t}{\dd t},
\end{equation}
where $m_t$ is the turnoff mass at time $t$ taken from \citet{Portinari98}.
For Pop I/II stellar IMF, we use
\begin{equation}
\Phi _{\rm PopI/II} (m) \propto  m^{-2.35} \exp \left(-\frac{m_{\rm ch}}{m} \right) 
\end{equation}
with $(m_{\min},~m_{\max},~m_{\rm ch}) = (0.1,~300,~0.35)~\Msun$.

The third and forth terms in the right hand side of Eq. (\ref{eq:dotMgas})
represent the interaction between a halo and the intergalactic medium (IGM).
The gas inflow rate from the IGM to a halo is
\begin{equation}
\frac{\dd M_{\rm inf}}{\dd t} = f_{\rm bar} \frac{\Mhalo}{t_{\rm acc}} ,
\end{equation}
where $f_{\rm bar} = \Omega _{\rm b} / \Omega _{\rm m}$ is the
baryon fraction.
A fraction of gas escapes from halo to IGM through SN explosions
at a rate
\begin{equation}
\frac{\dd M_{\rm ej}}{\dd t} = \frac{ 2 \epsilon _{\rm w}}
{V_{\rm esc}^2}  \frac{\dd \langle E_{\rm SN} \rangle}{\dd t} ,
\end{equation}
where $V_{\rm esc} = \left( G \Mhalo / \Rhalo \right) ^{1/2}$ is 
the escape velocity of the halo, and
$\dd \langle E_{\rm SN} \rangle / \dd t$ is the heating rate
from SNe as
\begin{equation}
\frac{\dd \langle E_{\rm SN} \rangle }{\dd t} (t) =
\int _{0} ^{t}  \frac{\dd e_{\rm SN}}{\dd t} (t - t^{\prime}) R_{\rm SF} (t^{\prime}) \dd t^{\prime} .
\end{equation}
The energy release rate $\dd e_{\rm SN} / \dd t$ per unit stellar mass is calculated as
\begin{equation}
\frac{\dd e_{\rm SN}}{\dd t} (t) =
\frac{E_{\rm SN} (m_t) \Phi (m_t) }{\int _{m_{\min}} ^{m_{\max}} m \Phi (m) \dd m}
\frac{\dd m_t}{\dd t} ,
\end{equation}
with a fraction $\epsilon _{\rm w}=0.002$ of the energy converted into the kinetic energy of gas dispersed into IGM \citep{Salvadori08}.

In Eq. (\ref{eq:dotMmet}), we treat the increase in metal mass in a similar way as gas mass.
The first term in the right hand side represents
the loss of metals by astration $Z_{\rm halo} R_{\rm SF}$, where $Z_{\rm halo} = \Mmet / \Mgas$
at the time $t$.
The second term is the metal yield by SNe:
\begin{equation}
\frac{\dd Y}{\dd t} (t) = 
\int _{0} ^{t}  \frac{\dd f_{\rm met}}{\dd t} (t - t^{\prime}) R_{\rm SF} (t^{\prime}) \dd t^{\prime} ,
\end{equation}
where $\dd f_{\rm met} / \dd t$ is the mass ejection rate of metals per unit stellar mass defined by
\begin{equation}
\frac{\dd f_{\rm met}}{\dd t} (t) = 
\frac{m_{\rm met} (m_t) \Phi (m_t)}{\int _{m_{\min}} ^{m_{\max}} m \Phi (m) \dd m}
\frac{\dd m_t}{\dd t} .
\end{equation}
In the third term, we estimate the inflow rate of metals from IGM to the halo to be
$Z_{\rm IGM} \dd M_{\rm inf} / \dd t$.
Here, we do not follow the metallicity evolution in the IGM. 
Instead, we just set $Z_{\rm IGM} = 0.01 Z_{\rm halo}$, assuming some metal dilution in the IGM. 
Metals are lost from a halo by SN explosions
at a rate $Z_{\rm w} \dd M_{\rm ej} / \dd t$ (fourth term), with the assumption $Z_{\rm w} = Z_{\rm halo}$.

We use $m_{\rm ej}$, $E_{\rm SN}$ and $m_{\rm met}$ from {\sc Yields Table 2013} \citep{Nomoto13}, which contains the contribution not only of CCSNe but also of asymptotic giant branch (AGB) stars for progenitor masses of $m = 1$--$8~\Msun$.
In using it, we assume the case of fixed stellar metallicity at $Z = 0.001$, but the yields do not change significantly for
other metallicities.

\subsubsection{Radiation feedback}
The LW radiation field in a galaxy is composed of the external and internal radiation components.

We calculate the 
external LW radiation intensity $J_{{\rm LW},i}^{\rm ext}$ at halo $i$ by summing up the contribution from source halos $j$ at distance $D_{ij}$, as
\begin{eqnarray}
J_{{\rm LW}, i}^{\rm ext} = \sum _{j \neq i} 
\frac{f_{\rm esc}}{4 \pi} \frac{h\nu _{\rm LW}}{\Delta \nu _{\rm LW}} \frac{Q_{{\rm LW}, j}}{4 \pi D_{ij}^2} ,
\label{eq:J_LW_ext}
\end{eqnarray}
where $h\nu _{\rm LW} = 12.8$ eV is the mean LW photon energy, and $h \Delta \nu _{\rm LW} = 2.4$ eV is the LW band width.
For the escape fraction $f_{\rm esc}$ of LW photons, 
\citet{Schauer15} found $f_{\rm esc} \lesssim 0.6$ for minihalos from cosmological simulations when taking the LW photon redshifting into consideration. 
As we also consider more massive halos ($\gtrsim 10^7~\Msun$), for which the escape fraction should be lower, we set $f_{\rm esc}=0.3$ regardless of halo mass for simplicity. 

We estimate internal radiation intensity $J_{{\rm LW}, i}^{\rm int}$ in halo $i$ by
\begin{equation}
J_{{\rm LW}, i}^{\rm int} = 
\frac{1}{4 \pi} \frac{h\nu _{\rm LW}}{\Delta \nu _{\rm LW}} \frac{Q_{{\rm LW}, i}}{4 \pi R_{{\rm gal}, i}^2} ,
\label{eq:J_LW_int}
\end{equation}
where $R_{{\rm gal}, i}$ is the radius of a galaxy hosted by the halo. 
We assume $R_{{\rm gal},i} = 0.1 R_{{\rm vir},i}$ \citep{Kauffmann93, Kauffmann99}.

The LW photon emissivity $Q_{{\rm LW},i}$ from halo $i$ is calculated by summing the contribution from Pop III ($Q_{\rm LW, III}$) and Pop II stars ($Q_{\rm LW, II}$).
The former is taken from
\citet{Schaerer02}, and 
the latter is calculated from the LW photon emissivity per unit stellar mass at age $t$ of a star cluster $q_{\rm LW, II} (t)$ taken from a population synthesis model {\sc starburst99}
\citep{Leitherer99} as
\begin{equation}
Q_{{\rm LW, II}} (t) =
\int _{t^{\prime}} ^{t} q_{\rm LW, II} (t - t^{\prime})  R_{\rm SF} (t^{\prime}) \dd t^{\prime}.
\end{equation}

\subsection{Criterion for DCBH formation}
\label{sec:DCcriterion}

In previous works including \citetalias{Chon16}, only chemically pristine halos are considered as candidate sites of SMS/DCBH formation 
 under the assumption that vigorous fragmentation of a star-forming cloud by efficient dust cooling prohibits SMS formation
(\citealt{Omukai08}, see however \citealt{Sassano21}).
Recently \citet{Chon20} demonstrated that, even with a finite amount of metals and dust grains and despite fragmentation induced by their cooling, a supermassive star can form via the super-competitive accretion at the center of a forming stellar cluster, as long as a sufficiently strong LW radiation field is present. 
They estimated the threshold metallicity $\Zth$ above which SMS formation is prohibited to be $10^{-4}$--$10^{-3}~\Zsun$, using Milky-Way like dust properties and dust-to-metal mass ratio ($\sim 0.5$). 

Considering the uncertainties in the condition for cloud fragmentation as well as in the dust amount in the early universe \citep{Schneider06}, we here leave $\Zth$ as a free parameter and explore four different values of $0$, $10^{-5}$, $10^{-4}$ and $10^{-3}~\Zsun$. 

The value of the critical LW intensity $\Jcr$ depends on the spectral energy distribution (SED) of the sources \citep{Omukai01, Shang10} because not only the LW photons but also infrared photons with $\gtrsim 0.75$ eV can suppress H$_2$ formation by destroying H$^-$, which is an intermediate product of the formation reaction. 
\citetalias{Chon16} used $\Jcr = 10^4$ and $10^2$ for radiation from Pop III and II sources, respectively, following the simulation result of \citet{Shang10}.
\citet{Sugimura14} found that $\Jcr \sim 10^3$, insensitive to stellar metallicity, by employing realistic SEDs of stellar populations at ages of $\lesssim 100$ Myr. 
Here we use a critical LW intensity of $\Jcr = 10^3$
for both Pop III and II sources.

We consider a sufficiently massive cloud hosted in an AC halo, which can gravitationally collapse despite H$_2$ dissociation by FUV irradiation.
A SMS/DCBH forms if the cloud has metallicity below the threshold value $< \Zth$ and is exposed to super-critical LW radiation $J_{21} > \Jcr$.
The cloud/halo should continue to satisfy this condition for a certain interval of time for the cloud to collapse and form a SMS. 
We here impose a minimum threshold time $\tth = 4$ Myr, corresponding to the free-fall time of a cloud with density $100~\percc$.

The criteria for DCBH formation can be summarized as follows:
\begin{enumerate}
\item a halo to be massive enough with the virial temperature $\Tvir > 8000$ K,
\item the radiation in the halo to be above the critical LW intensity $\Jcr = 10^3$,
\item the metallicity in the halo to be below the threshold metallicity $\Zth$ and
\item the above criteria to be satisfied for at least $\tth = 4$ Myr. 
\end{enumerate}
All halos that meet the DC criteria ("DC halos") have been found to have sufficient gas content to form a seed BH of $10^5~\Msun$.
A progenitor halo may meet the criteria twice or more before merging with another halo. 
In such a case, considering that most of the gas might have been consumed in the first DCBH formation, we assume that only one DCBH forms in the halo.

Here, we study the impact of internal radiation on the number of formed DCBHs, but we also perform conparison runs without internal radiation. 
We call the runs with and without the internal radiation as {\SRi} and {\SRo}, respectively.
In each run, we perform the calculations for four different threshold metallicities of
$\Zth = 0$, $10^{-5}$, $10^{-4}$ and $10^{-3}~\Zsun$. 

\begin{figure*}
\includegraphics[width=\textwidth]{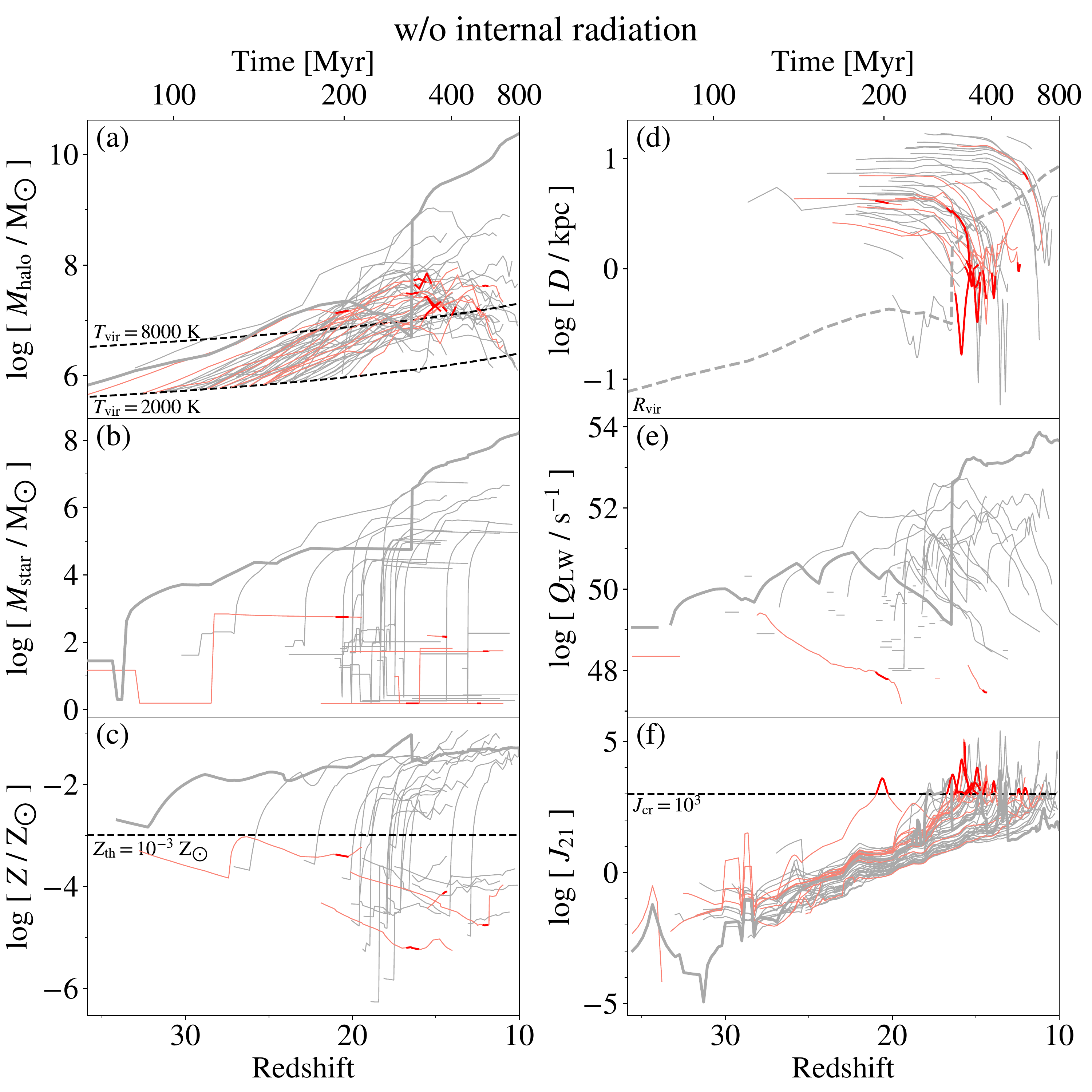}
\caption{Time evolution of halo physical properties in a run without internal radiation ({\SRo}).
We plot the result for a halo that host the first Pop III star (``main halo''; thick curves)
and its progenitor satellites (thin curves) in the most overdense region {\tt H1}.
The panels show (a) the halo mass, (b) stellar mass, (c) metallicity, (d) distance of the progenitors
from the main halo in physical coordinates, (e) LW emissivity from each halo and (f) LW intensity in each halo.
A progenitor merges with another halo at the redshift where the curve is truncated.
The curves are highlighted in red when the halos satisfy the
criterion for direct collapse (see Section \ref{sec:DCcriterion}) 
with a threshold metallicity $\Zth = 10^{-3}~\Zsun$.
The dashed lines in different panels indicate:
(a) the halo masses for the
virial temperatures of $2000$ and $8000$ K, (c) the threshold metallicity $\Zth = 10^{-3}~\Zsun$, (d) the virial radius of the main halo, and (f) the critical LW intensity $\Jcr = 10^3$.}
\label{fig:tZJ_SR0_data_4_0_0_609_7403490}
\end{figure*}

\begin{figure*}
\includegraphics[width=\textwidth]{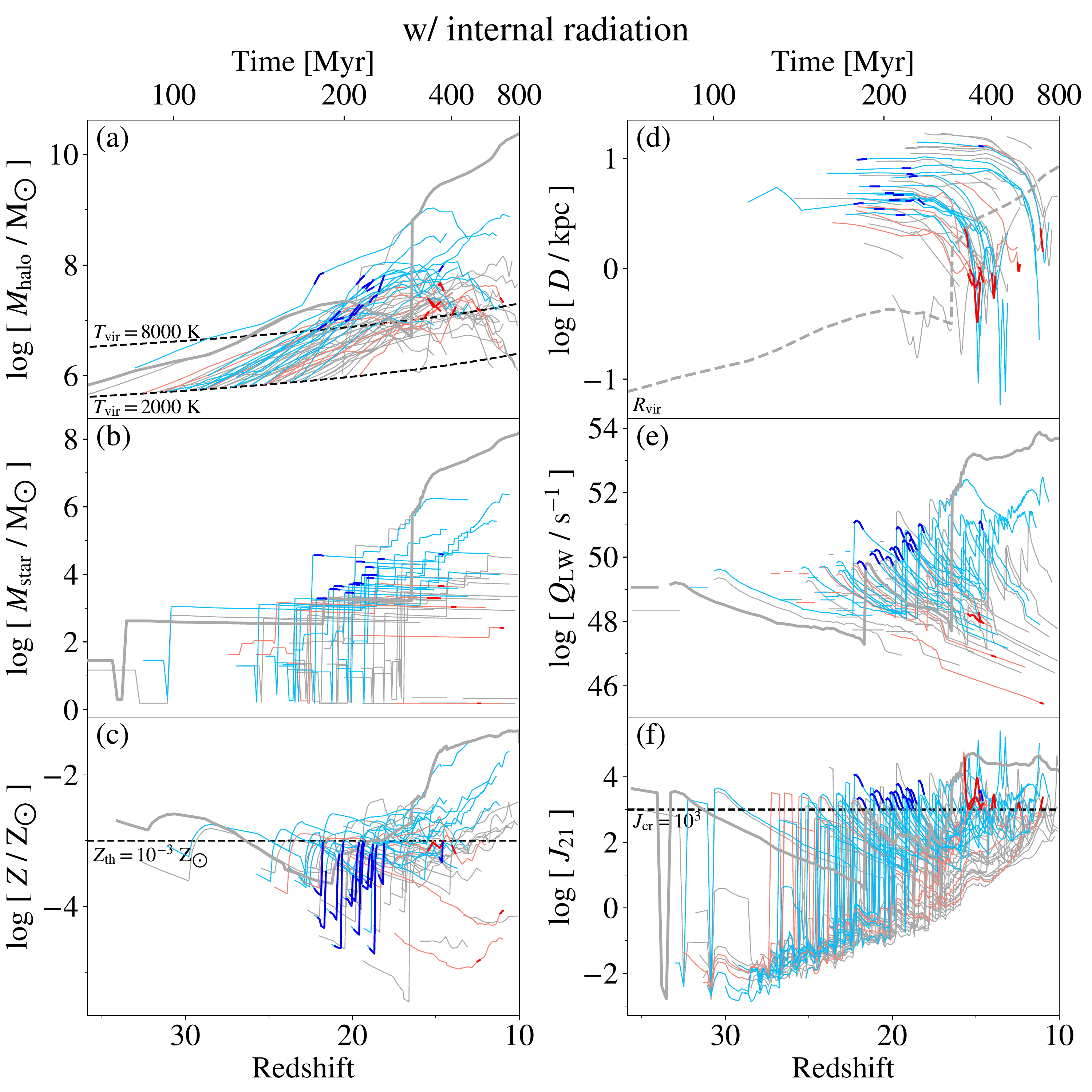}
\caption{Same as Fig. \ref{fig:tZJ_SR0_data_4_0_0_609_7403490}, but for the run with internal radiation ({\SRi}).
The curves are highlighted in red and blue in the redshift ranges where external and internal radiation induces direct collapse, respectively.}
\label{fig:tZJ_SR1_data_4_0_0_609_7403490}
\end{figure*}

\section{Results}
\label{sec:results}

In this section, we first present physical properties of halos with particular emphasis on the progenitors of the most massive halo at the final redshift $\zfin = 10$ in the most overdense region {\tt H1} (Section \ref{sec:evolution}).
We then estimate the number density $\nDCBH$ of DCBHs in each zoom-in region and discuss the dependence of $\nDCBH$ on whether the internal radiation is considered or not, as well as on the adopted values of the threshold metallicity
in Section \ref{sec:dependency_of_nDCBH}.
In Section \ref{sec:nDCBH}, we estimate the average number density of DCBHs in a cosmological volume.
Finally, we show the average number of DCBHs hosted by a halo as a function of halo mass
at $\zfin = 10$ in Section \ref{sec:barNdcbh_z10}.

\subsection{Evolution of metallicity and LW intensity in halos}
\label{sec:evolution}

In the most overdense region {\tt H1}, 5,866 merger trees are identified until the final redshift $\zfin = 10$.
Figs. \ref{fig:tZJ_SR0_data_4_0_0_609_7403490} and \ref{fig:tZJ_SR1_data_4_0_0_609_7403490} show the time evolution of the physical properties of the most massive halo and its progenitors for {\SRo} and {\SRi}, respectively.
The thick grey curves indicate a halo that hosts the first Pop III star 
(``main halo'').
The main halo is one of the most massive halos and the dominant LW source for its satellites in most of the merger history.
The panels show (a) the halo mass, (b) stellar mass, (c) metallicity, (d) distance of progenitors
from the main halo in physical coordinates, (e) LW emissivity from each halo and (f) LW intensity in each halo.
In this section, we first describe the result of the {\SRo} run as a reference 
case (Section \ref{sec:evolution_SRo}).
Then, we discuss the result of the {\SRi} run, comparing with the {\SRo} run
(Section \ref{sec:evolution_SRi}).

\subsubsection{Cases without internal radiation}
\label{sec:evolution_SRo}

Here we present the results of the {\SRo} run, which we consider our reference case.
First, we describe the evolution of the main halo.
The first star forms at redshift $z=36$ in the main halo when its virial temperature exceeds 2,000 K (lower dashed curve in Fig. \ref{fig:tZJ_SR0_data_4_0_0_609_7403490}a).
The star explodes as a SN at $z=33$
(Fig. \ref{fig:tZJ_SR0_data_4_0_0_609_7403490}b).
This causes a sudden decline in $\Mstar$ from the main-sequence mass to the remnant mass.
At the same time, the metallicity jumps to $\sim 10^{-3}~\Zsun$ by SN metal injection. 
Then, the second-generation of stars form as Pop II stars, and $\Mstar$ increases abruptly to $\sim 500~\Msun$, immediately after this event.
Thanks to repeated star formation episodes, the stellar mass, along with the metallicity, gradually increases
(Fig. \ref{fig:tZJ_SR0_data_4_0_0_609_7403490}b and c)
until tidal interaction with a more massive halo causes a loss of DM mass below $z\simeq 20$
(Fig. \ref{fig:tZJ_SR0_data_4_0_0_609_7403490}a). 
The two halos merge at $z=17$, inducing a sudden increase by two orders of magnitude of both the halo and stellar masses.
The main halo continues to grow in mass by mergers with other satellites, and the virial radius increases (grey dashed curve in Fig. \ref{fig:tZJ_SR0_data_4_0_0_609_7403490}d).
The stellar mass also increases as a result of star formation sustained by gas accretion from the IGM. 
In contrast, the metallicity experiences only a mild increase, because the metal ejection from massive stars into the ISM is accompanied by pristine-gas accretion into the halo.
The LW photon emissivity increases with stellar mass
(Fig. \ref{fig:tZJ_SR0_data_4_0_0_609_7403490}e), resulting in enhanced LW irradiation onto neighboring satellites.

Next, we present the evolution of the metallicity and LW intensity in satellite halos.
The UV intensity $J_{21}$ in a satellite increases almost monotonically as it approaches the main halo (thin curves in Fig. \ref{fig:tZJ_SR0_data_4_0_0_609_7403490}d). 
About an order of magnitude scatter in $J_{21}$ among the satellites causes different star formation histories (Fig. \ref{fig:tZJ_SR0_data_4_0_0_609_7403490}b).
In some halos, strong LW irradiation (indicated by the red curves) suppresses star formation and $\Mstar$ is smaller  ($\lesssim 10^3~\Msun$).
The metallicity in such halos remains low due to fewer SN events (Fig. \ref{fig:tZJ_SR0_data_4_0_0_609_7403490}c).
Later, just before merging with the main halo,
thanks to the decreasing distance to the main halo (Fig. \ref{fig:tZJ_SR0_data_4_0_0_609_7403490}d),
the LW intensity jumps up to
\begin{equation}
J_{21}^{\rm ext} = 7 \times 10^2 
\left( \frac{f_{\rm esc}}{0.3} \right)
\left( \frac{Q_{\rm LW}}{10^{53}~{\rm s}^{-1}} \right)
\left( \frac{D}{1~{\rm kpc}} \right)^{-2}
\label{eq:J21ext}
\end{equation}
as seen in Fig. \ref{fig:tZJ_SR0_data_4_0_0_609_7403490}f (Eq. \ref{eq:J_LW_ext}), exceeding the DC critical value.
The halos eventually merge with the main halo after some inspiral motions.
The inspiral timescale can be estimated as the dynamical time of the main halo:
\begin{equation}
t_{\rm dyn} = \left( \frac{3\pi}{32 G \rho _{\rm vir}} \right)^{1/2} \simeq 45.1~{\rm Myr}
\left( \frac{1+z}{16} \right)^{-3/2},
\end{equation}
where $\rho _{\rm vir} = 200 \bar \rho$ is the average density of halos.
The timescale is longer than the threshold time $\tth=4$ Myr.
This means that the DC criterion can be satisfied in some of those halos during the inspiral:
11 out of 52 ACHs (red curves) satisfy the condition for $\Zth = 10^{-3}~\Zsun$ in the merger tree of the most massive halo in {\tt H1},
in the redshift intervals
highlighted by the thick red curves in Fig. \ref{fig:tZJ_SR0_data_4_0_0_609_7403490}.

\begin{table}
\begin{minipage}{\columnwidth}
\caption{Number density of ACHs and DCBHs at $z=10$}
\label{tab:nDCBH}
\begin{tabular}{ccccccc}
\hline
Run & $\nACH$ [$\Mpc ^{-3}$] & \multicolumn{4}{c}{$\nDCBH$ [$\Mpc^{-3}$]} & $f_{\rm int}$ \\
    & $\Zcr$ [$\Zsun$] & $0$ &  $10^{-5}$  & $10^{-4}$              & $10^{-3}$ & \\
\hline
\multicolumn{6}{c}{{w/o internal radiation}} \\
\hline
{\tt H1} & $146.$ & $5.55$ & $6.20$ & $8.33$ & $9.47$ &        \\
{\tt H2} & $111.$ & $3.75$ & $4.24$ & $6.20$ & $7.35$ &        \\
{\tt H3} & $79.5$ & $1.96$ & $2.29$ & $3.26$ & $3.92$ &        \\
{\tt H4} & $83.7$ & $2.45$ & $3.10$ & $3.75$ & $4.24$ &        \\
{\tt M1} & $62.0$ & $1.31$ & $1.31$ & $1.31$ & $1.47$ &        \\
{\tt M2} & $71.8$ & $2.45$ & $2.61$ & $3.43$ & $3.92$ &        \\
{\tt M3} & $63.0$ & $0.98$ & $0.98$ & $1.31$ & $1.63$ &        \\
{\tt M4} & $69.4$ & $0.65$ & $0.65$ & $0.98$ & $1.31$ &        \\
{\tt L1} & $33.1$ & $0.16$ & $0.33$ & $0.33$ & $0.65$ &        \\
{\tt L2} & $19.3$ & $0.16$ & $0.16$ & $0.16$ & $0.16$ &        \\
{\tt L3} & $18.3$ & $0.00$ & $0.00$ & $0.00$ & $0.16$ &        \\
{\tt L4} & $22.2$ & $0.16$ & $0.16$ & $0.16$ & $0.33$ &        \\
Average  & $17.0$ & $0.20$ & $0.23$ & $0.28$ & $0.36$ &        \\
\hline
\multicolumn{6}{c}{w/ internal radiation} \\
\hline
{\tt H1} & $146.$ & $1.14$ & $1.14$ & $1.80$ & $19.9$ & $0.77$ \\
{\tt H2} & $111.$ & $0.98$ & $0.98$ & $2.29$ & $14.9$ & $0.76$ \\
{\tt H3} & $79.5$ & $0.49$ & $0.65$ & $0.65$ & $11.9$ & $0.90$ \\
{\tt H4} & $83.7$ & $0.82$ & $0.82$ & $0.98$ & $13.2$ & $0.84$ \\
{\tt M1} & $62.0$ & $0.00$ & $0.00$ & $0.00$ & $3.43$ & $1.00$ \\
{\tt M2} & $71.8$ & $0.16$ & $0.16$ & $0.82$ & $6.86$ & $0.69$ \\
{\tt M3} & $63.0$ & $0.00$ & $0.00$ & $0.00$ & $3.10$ & $0.95$ \\
{\tt M4} & $69.4$ & $0.00$ & $0.00$ & $0.00$ & $3.92$ & $0.87$ \\
{\tt L1} & $33.1$ & $0.16$ & $0.16$ & $0.16$ & $1.80$ & $0.73$ \\
{\tt L2} & $19.3$ & $0.00$ & $0.00$ & $0.00$ & $0.82$ & $1.00$ \\
{\tt L3} & $18.3$ & $0.00$ & $0.00$ & $0.00$ & $1.14$ & $0.86$ \\
{\tt L4} & $22.2$ & $0.00$ & $0.00$ & $0.00$ & $0.65$ & $1.00$ \\
Average  & $17.0$ & $0.03$ & $0.03$ & $0.06$ & $1.01$ & $0.89$ \\
\hline
\end{tabular}
\medskip \\
Note --- 
(1) ID of zoom-in regions.
(2) number density $\nACH$ of ACHs.
(3--6) number density $\nDCBH$ of DCBHs with threshold metallicties $\Zth=0$, $10^{-5}$, $10^{-4}$ and $10^{-3}~\Zsun$.
(7) fraction $f_{\rm int}$ of DCBHs that form due to internal radiation in the case of $\Zth = 10^{-3}~\Zsun$.
The fraction is zero for the other runs.
\end{minipage}
\end{table}

\begin{figure}
\includegraphics[width=\columnwidth]{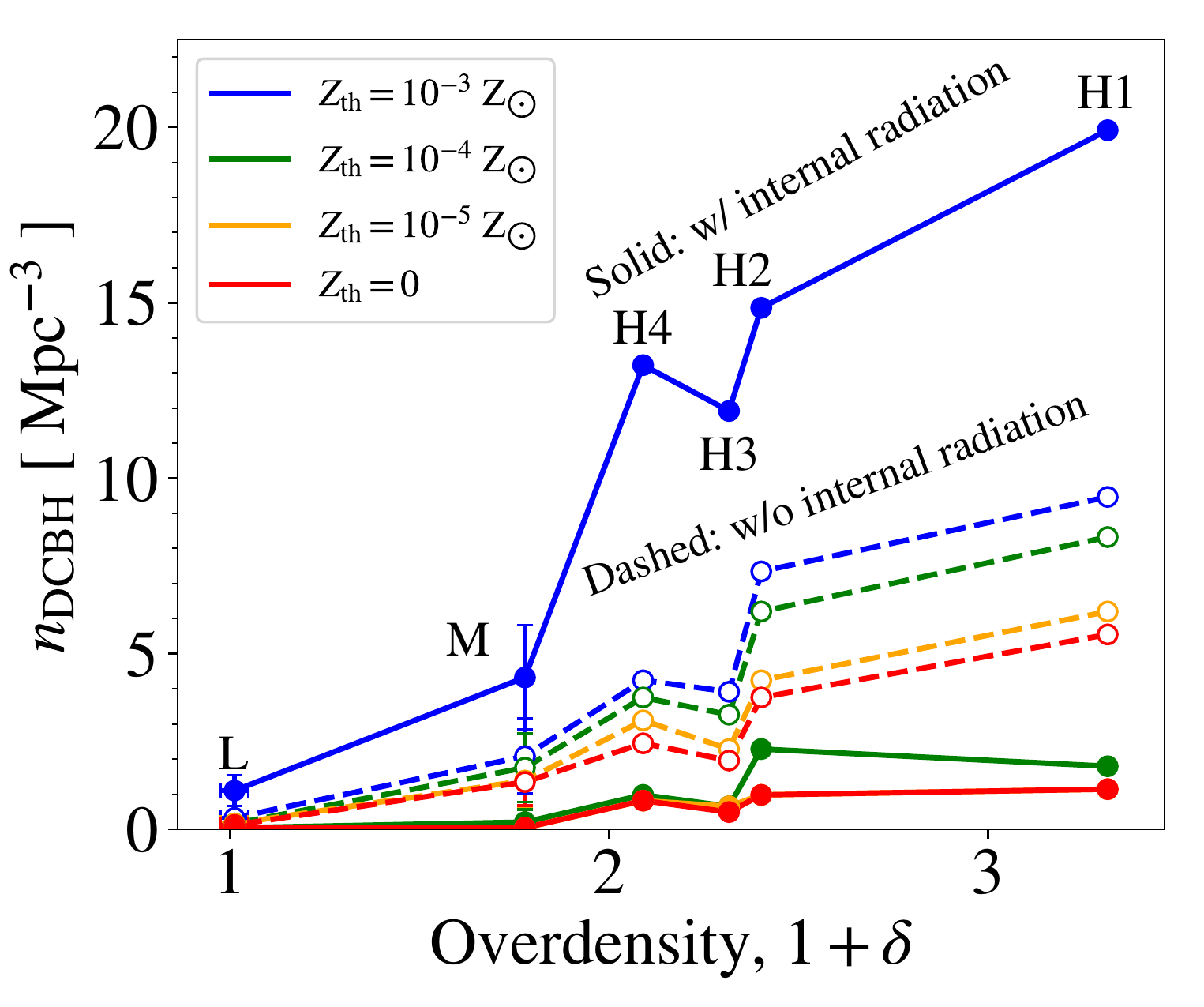}
\caption{Number density $\nDCBH$ of DCBHs as a function of overdensity, $1+\delta$, of each zoom-in region.
The dashed and solid curves show the results 
in the cases without and with internal radiation, respectively.
The red, orange, green and blue curves show the results with threshold metallicities $\Zth = 0$, $10^{-5}$, $10^{-4}$ and $10^{-3}~\Zsun$, respectively.
Since $1+\delta$ and $\nDCBH$ of {\tt L1}-{\tt 4} and {\tt M1}--{\tt 4} are similar within the error bars, we show the average values as indicated by ``L'' and ``M'', respectively.}
\label{fig:OD-nDCBH}
\end{figure}

\begin{figure*}
\includegraphics[width=\textwidth]{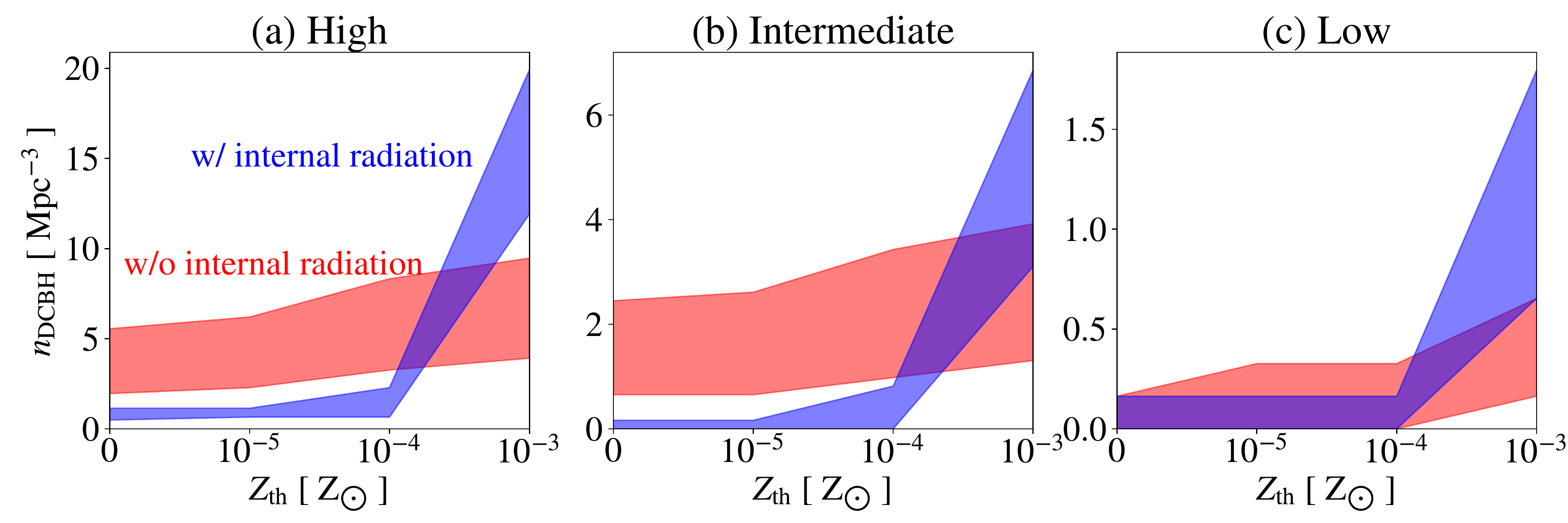}
\caption{The number density $\nDCBH$ of halos that host DCBHs
as a function of threshold metallicity $\Zth$.
In each plot the range of $\nDCBH$ in the four zoom-in regions with (a) high, (b) intermediate and (c) low overdensities is indicated.
The red and blue shaded regions denote the results without and with internal radiation, respectively.}
\label{fig:nDCBH}
\end{figure*}

\subsubsection{Cases with internal radiation}
\label{sec:evolution_SRi}
Next, we present the result in the run {\SRi} and compare it with the result in the run {\SRo}.
Inclusion of internal radiation not only affects the LW intensity but also modifies the star formation history in each halo.
First, we describe the evolution of the main halo (grey thick curves in Fig. \ref{fig:tZJ_SR1_data_4_0_0_609_7403490}).
Until the first star forms,
the evolution is almost identical to that in the {\SRo} run. 
The inclusion of internal radiation, however, causes a different evolution of $\Mstar$ 
after the second-generation stars form.
For {\SRo}, $\Mstar$ continues to increase gradually (Fig. \ref{fig:tZJ_SR0_data_4_0_0_609_7403490}b).
On the other hand, for {\SRi}, star formation is suspended, and $\Mstar$ remains almost constant for a few$\times 10$ Myr (Fig. \ref{fig:tZJ_SR1_data_4_0_0_609_7403490}b) 
due to strong internal LW radiation.
At this moment, the LW intensity can be estimated as
\begin{equation}
J_{21}^{\rm int} = 940
\left( \frac{Q_{\rm LW}}{10^{50}~{\rm s}^{-1}} \right)
\left( \frac{R_{\rm vir}}{0.5~{\rm kpc}} \right)^{-2}
\end{equation}
(Fig. \ref{fig:tZJ_SR0_data_4_0_0_609_7403490}e and f).
When massive stars start to die, 
$Q_{\rm LW}$ and $J_{21}$ decrease slowly.  
$J_{21}$ becomes as small as $\sim 1$ at $z=22$, and a third episode of star formation 
occurs and increases the stellar mass $\Mstar$ ($\sim 3000~\Msun$) as well as $Q_{\rm LW}$.
Sudden increase in $\Mstar$ at $z=17$ corresponds to the merger with another more massive halo.
The stellar mass and metallicity continue to increase further after this event.

Internal radiation can modify the evolution of the satellites (thin curves).
Star formation is intermittent also in the satellites
as seen in step-wise increase of the stellar mass $\Mstar$. The starburst and quiescent phases alternate with a period of a few$\times 10$ Myr (Fig. \ref{fig:tZJ_SR1_data_4_0_0_609_7403490}b) .
Rapid increase of $J_{21}$ due to starbursts prevents further star formation, followed by slow decrease of $J_{21}$ as massive stars die.
Less active star formation in the satellites results in smaller metallicities ($\sim 10^{-3}~\Zsun$) than for {\SRo} (Fig. \ref{fig:tZJ_SR1_data_4_0_0_609_7403490}c).
Star formation proceeds in such a self-regulated manner until the external radiation becomes dominant when the satellites enter within the virial radius of the main halo
(grey dashed curve in \ref{fig:tZJ_SR1_data_4_0_0_609_7403490}d).

In the {\SRi} case, both external and internal radiation contributes to 
the LW intensity in the satellites.
Out of 52 ACHs in the simulation, 14 and 7 ACHs satisfy the DC criterion
thanks to internal and external radiation, respectively, as indicated by
the blue and red thick curves in Fig. 
\ref{fig:tZJ_SR1_data_4_0_0_609_7403490}. 
This means that the internal radiation triples the number of DC halos.
If the metallicity in a satellite is below the threshold ($\lesssim 10^{-3}~\Zsun$), 
radiation from the main halo increases as they get closer and exceeds $\Jcr$ just before their merger (Fig. \ref{fig:tZJ_SR1_data_4_0_0_609_7403490}f), thereby inducing the DC (red curves in Fig. \ref{fig:tZJ_SR1_data_4_0_0_609_7403490}d), as in the {\SRo} case.

Halos that rapidly grow in mass to exceed $\Tvir \geq 8000$ K meet
the DC criterion thanks to internal radiation (blue curves in Fig. \ref{fig:tZJ_SR1_data_4_0_0_609_7403490}).
This occurs at the first episode of Pop II star formation as indicated by the first rapid increase of $\Mstar$ in Fig. \ref{fig:tZJ_SR1_data_4_0_0_609_7403490}b.
At this moment, the metallicity is still small, $Z \sim 10^{-4}~\Zsun$, because only Pop III SNe have enriched the halos.
Additionally, the internal LW intensity can reach
\begin{equation}
J_{21}^{\rm int} = 2.35\E{3} 
\left( \frac{Q_{\rm LW}}{10^{51}~{\rm s}^{-1}} \right)
\left( \frac{D}{0.1~{\rm kpc}} \right)^{-2}
\label{eq:J21int}
\end{equation}
(eq. \ref{eq:J_LW_int}) only after a single star formation event (Fig. \ref{fig:tZJ_SR1_data_4_0_0_609_7403490}f).
This condition lasts for $\sim 4$ Myr,
until massive stars die as SNe and no longer emit LW photons.
Since the time duration is comparable to the threshold time $\tth$ set in our criterion (iv), the DC criterion is satisfied in ACHs under our parameter choice of $\Zth = 10^{-3}~\Zsun$ and $\tth = 4$ Myr.

It should also be noted that internal radiation may also promote DCBH formation in the main halo of each merger tree.
Without internal radiation, DCBH formation occurs mostly in satellite halos synchronized with a star formation activity of the nearby main halo \citep{Agarwal12, Chon16}.
In the presence of internal radiation, the fraction of merger trees where the DCBH can form in the main halo increases from 44\% to 68\% in {\tt H1} with $\Zth = 10^{-3}~\Zsun$.
This fraction increases by a factor of two in the other cases as well.

\subsection{Dependence of DCBH formation rate on overdensity,
threshold metallicity and internal radiation}
\label{sec:dependency_of_nDCBH}

We summarize the results in all the runs we studied in Table \ref{tab:nDCBH}.
The table reports the number density of ACHs (2nd column) and DCBHs for the cases with different threshold metallicities (3--6th column) in each zoom-in region.
We derive the number density by dividing the number of ACHs and DCBHs at $z=10$ by the comoving volume $\left( 1.24 h^{-1}~\Mpc \right) ^3 = 6.12~\Mpc ^3$ of each zoom-in region.
The results demonstrate that $\nDCBH$ depends on overdensity and threshold metallicity, and it is affected by the inclusion of internal radiation.
In this section, we discuss those effects in this order.

\begin{figure}
\includegraphics[width=0.95\columnwidth]{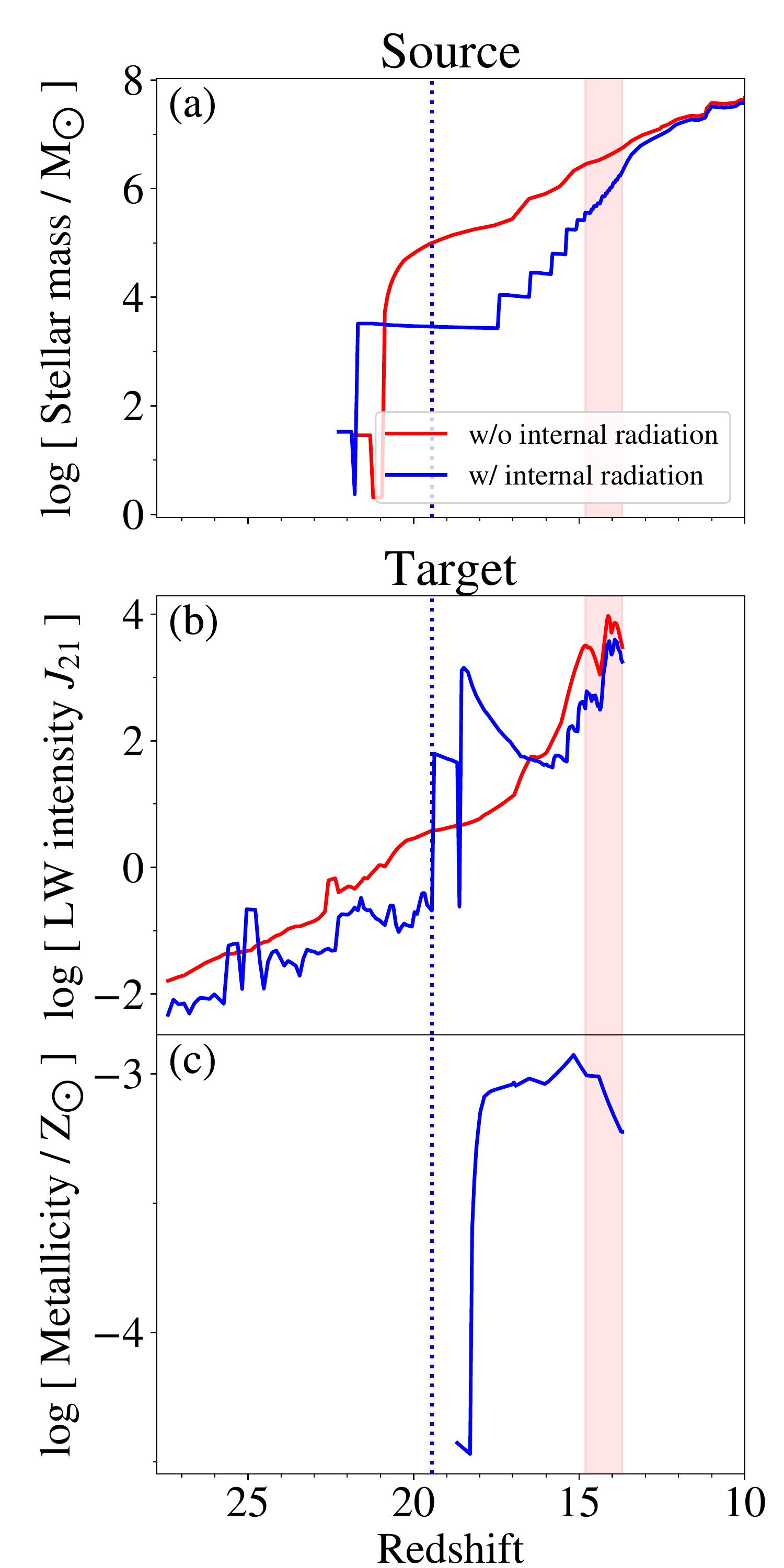}
\caption{Comparison of physical properties of a LW source halo and
a target halo in the most overdense zoom-in region {\tt H1} as a function of redshift.
The panels show 
(a) stellar mass of the source and
(b) LW intensity and
(c) metallicity of the target 
in the runs without (red curves) and with (blue curves) internal radiation.
The blue dotted line indicates the redshift of the first star
formation in the case with internal radiation.
The red shaded region shows the redshift range where our
criteria for DCBH formation are satisfied in the case without internal radiation
for threshold metallicity $\Zth = 0$.
The metallicity in the target is zero in the case without internal radiation.
The target merges with the source at redshift $13.7$.}
\label{fig:tZJcmp_data_4_0_0_609}
\end{figure}

\subsubsection{Overdensity}
Here, we see the effect of overdensity on DCBH formation.
Fig. \ref{fig:OD-nDCBH} shows the number density $\nDCBH$ as a function of overdensity in each zoom-in region.
We show the average values of $\nDCBH$ and $\delta$ among {\tt L1}--{\tt 4} (``L'') and {\tt M1}--{\tt 4} (``M''), because these values are similar within the error bars. 
We plot the results in the cases of {\SRo} (dashed) and {\SRi} (solid).
The red, orange, green and blue curves show the cases with threshold metallicities $\Zth = 0$, $10^{-5}$, $10^{-4}$ and $10^{-3}~\Zsun$, respectively.

The number density of DCBHs increases with increasing overdensity of the zoom-in regions for any threshold value of metallicity (Fig. \ref{fig:OD-nDCBH}).
This is attributable to larger number density $\nACH$ of ACHs in more overdense regions (Table \ref{tab:nDCBH}).
Their mutual separation also becomes smaller and the LW intensity in each halo is enhanced due to the external radiation. 

Although the number density of DCBHs has been predicted by earlier works,
its dependence on overdensity has not been investigated in detail \citep{Agarwal12, Johnson13, Habouzit16, Chon16, Li21, Spinoso22}.
In this work, we comprehensively study DCBH formation not only in the overdense regions ($3$--$4\sigma$) but also in the less overdense regions down to $\sim 0\sigma$ (Table \ref{tab:zoomin}) for the first time.
Although $\nDCBH$ is smaller in less dense regions, we find some DCBHs still form even in average density ($\sim 0\sigma$) regions (Table \ref{tab:nDCBH}) with important implications for the demography of nuclear black holes as we will discuss later.

\subsubsection{Threshold metallicity}
\label{sec:effect_of_threshold_metallicity}
We allowed DCBH formation not only in metal-free environments but also in regions with metallicity below a threshold value $\Zth$.
The variations in the number of DCBH forming in the different runs provide an indication of the uncertainties related  to the value of $\Zth$.
Fig. \ref{fig:nDCBH} shows the DCBH number density $\nDCBH$ as a function of $\Zth$. 
We plot the range of $\nDCBH$ in the four zoom-in regions with (a) high, (b) intermediate and (c) low overdensities, for the cases with ({\SRi}, blue) and without ({\SRo}, red) internal radiation, respectively.
For larger threshold metallicity, 
more DCBHs are formed in each zoom-in region, as expected.
However, the behaviour of $\nDCBH$ with increasing $Z_{\rm th}$ is very different when we compare the {\SRo} and {\SRi} cases.
In models without internal radiation ({\SRo}), $\nDCBH$ does not depend strongly on $\Zth$, increasing only by a factor of two when $\Zth$ increases from $0$ to $10^{-3}~\Zsun$ (red shaded regions in Fig. \ref{fig:nDCBH}).
The reason is that
those DC halos are almost continuously exposed to strong LW radiation (red curves in Fig. \ref{fig:tZJ_SR0_data_4_0_0_609_7403490}f) and their metallicities remain much lower than $10^{-3}~\Zsun$ (Fig. \ref{fig:tZJ_SR0_data_4_0_0_609_7403490}c).
Since the gas is almost metal-free when the DCBH condition is satisfied, raising $\Zth$ from 0 to some finite value does not significantly affect the number of DCBH halos $\nDCBH$.
On the other hand, in the cases with internal radiation ({\SRi}), $\nDCBH$ increases only mildly at $\Zth \lesssim 10^{-4}~\Zsun$, but jumps up by an order of magnitude between $10^{-4}$ and $10^{-3}~\Zsun$ (the blue shaded regions in Fig. \ref{fig:nDCBH}).
In most of halos with internal LW intensities above $\Jcr$, the mass of stars (LW and metal sources) and accordingly the metallicity are large ($\gtrsim 10^3~\Msun$ and $\gtrsim 10^{-4}~\Zsun$, respectively).
These halos satisfy the DC criterion only when $\Zth$ is as high as $10^{-3}~\Zsun$.
Then, changing the adopted value of $\Zth$ from $10^{-4}$ to $10^{-3}~\Zsun$ 
makes a big difference on $\nDCBH$.
In fact, while $70$--$100\%$ of DCBH are allowed to form due to internal radiation in the {\SRi} run with $\Zth=10^{-3}~\Zsun$ (the seventh column in Table \ref{tab:nDCBH}), DCBH formation is entirely due to external radiation in the runs with lower $\Zth$.

\subsubsection{Internal radiation}
\label{sec:effect_of_internal_radiation}
Naively, we expect that the addition of internal radiation enhances the number of DCBHs.
In reality, however, $\nDCBH$ is smaller for {\SRi} than for {\SRo} in the case of $\Zth \leq 10^{-4}~\Zsun$:
some halos satisfy the DC criterion in {\SRo}, but not in {\SRi} run.
To see the reason for this, we compare the evolution of a representative halo in the {\SRo} and {\SRi} runs for $\Zth = 0$ (Fig. \ref{fig:tZJcmp_data_4_0_0_609}).
Since external LW radiation is dominant in this halo (hereafter called ``target''), we analyze the properties of both the target halo and the LW source halo.
In Fig. \ref{fig:tZJcmp_data_4_0_0_609}, we plot 
(a) the stellar mass in the source and
(b) the LW intensity and (c) metallicity in the target as a function of redshift.
The stellar mass in the source 
is smaller in {\SRi} (blue curve) than in {\SRo} (red curve) at $z<22$ due to the suppression of star formation by internal radiation
(Fig. \ref{fig:tZJcmp_data_4_0_0_609}a).
Accordingly, the LW emissivity from the source, as well as the intensity of the radiation field on the target, is smaller before the first star formation 
(Fig. \ref{fig:tZJcmp_data_4_0_0_609}b).  
With the weaker radiation field,
star formation occurs in the target in {\SRi} run at $z=19$ (dotted vertical line), while it does not occur in the {\SRo} run.
After a single SN event,
the metallicity jumps up to $\sim 10^{-3}~\Zsun$ (Fig. \ref{fig:tZJcmp_data_4_0_0_609}c).
Consequently, DC is prohibited in {\SRi} run for $\Zth \leq 10^{-4}~\Zsun$. 
Only with $\Zth$ as high as $10^{-3}~\Zsun$, the number of DCBH becomes larger for {\SRi} than for {\SRo} as a result of larger LW intensity due to internal radiation, as discussed in the previous section.

\section{Estimate of DCBH number density}
\label{sec:Ndcbh}

In this section, we estimate statistical properties of DCBHs formed in our model and compare them to observations.
We first derive the global average of DCBH number density (Section \ref{sec:nDCBH}).
Then, we present the halo occupation fraction of DCBHs as a function of halo mass at $\zfin = 10$ (Section \ref{sec:barNdcbh_z10}) and extrapolate this to the local Universe (Section \ref{sec:barNdcbh_z0}).

\subsection{Global average of DCBH number density}
\label{sec:nDCBH}
From the overdensity distribution function (Eq. \ref{eq:od}) and
the number density of DCBHs in each zoom-in region (Table \ref{tab:nDCBH}),
we compute the global average of $\nDCBH$ in the entire cosmological volume.
For each threshold metallicity, both with and without internal radiation,
$\nDCBH$ is almost proportional to the overdensity (Fig. \ref{fig:OD-nDCBH})
and can be fit with a linear function.
The convolution of the linear function with Eq. \ref{eq:od} gives the global average $\barnDCBH$
as indicated in the rows ``Average'' of Table \ref{tab:nDCBH}.
When $\Zth < 10^{-3}~\Zsun$, 
$\barnDCBH$ for {\SRi} ($0.03$--$0.06~\Mpc ^{-3}$) is smaller than $\barnDCBH$ for the {\SRo} model ($0.20$--$0.28~\Mpc ^{-3}$), for the reasons discussed in Section \ref{sec:effect_of_internal_radiation}:
star formation in LW source halos is suppressed by their own internal radiation.
With increasing threshold metallicity, $\barnDCBH$ increases.
When $\Zth = 10^{-3}~\Zsun$, $\barnDCBH$ is estimated to be $1.0~\Mpc ^{-3}$ in the {\SRi} model, which is larger than $0.36~\Mpc ^{-3}$ in the {\SRo} model.
This indicates that $\barnDCBH$ is underestimated by a factor of three if the effect of internal radiation is not considered.

\begin{figure}
\includegraphics[width=\columnwidth]{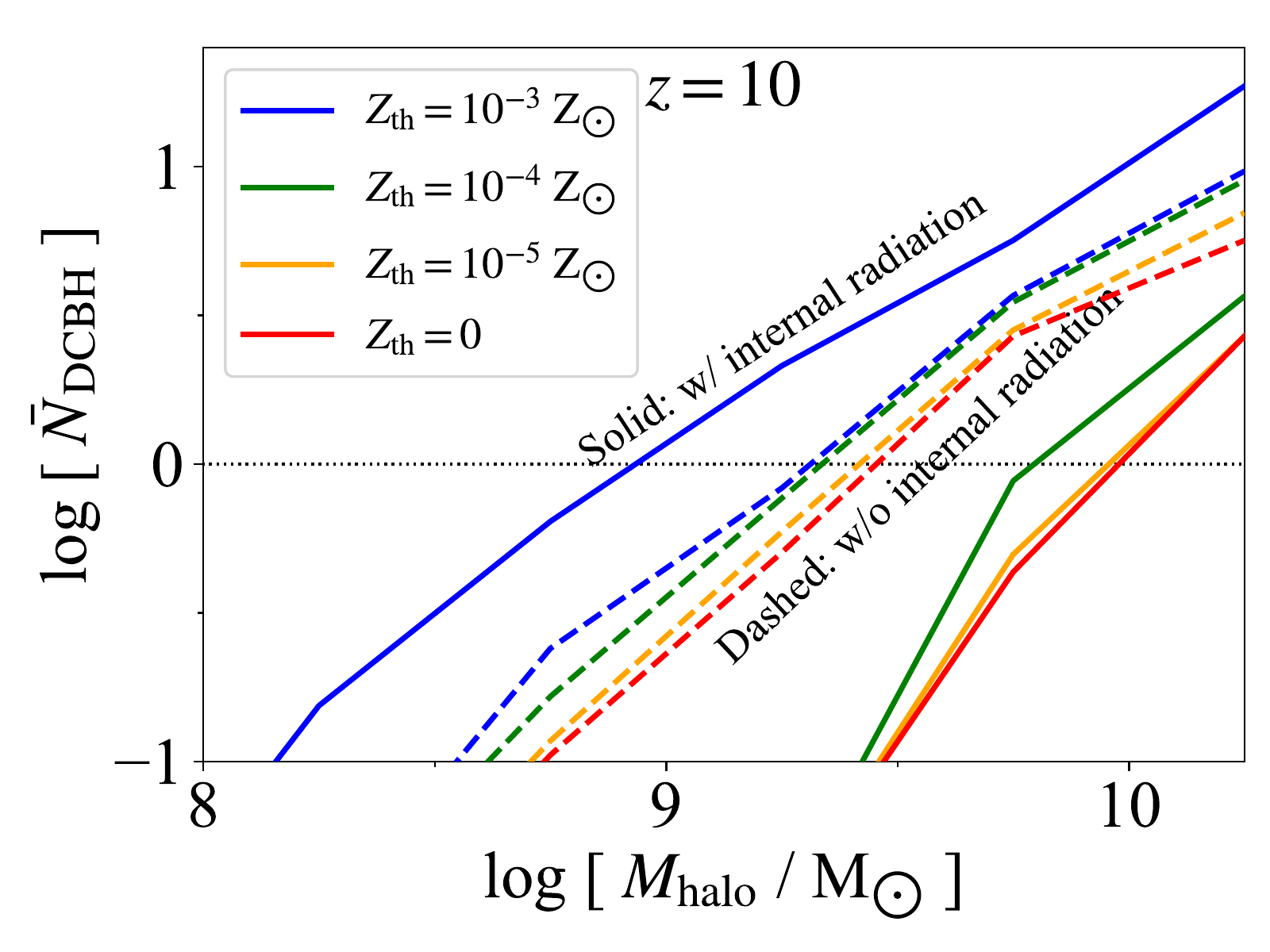}
\caption{Average number $\bar N_{\rm DCBH}$ of DCBHs per halo
as a function of halo mass $\Mhalo$ at redshift $z=10$.
The dashed and solid curves indicate the cases without and
with internal radiation, respectively.
The red, orange, green and blue curves indicate the cases with
threshold metallicities $\Zth = 0$, $10^{-5}$, $10^{-4}$ and $10^{-3}~\Zsun$, respectively.
The horizontal dotted curve indicates $\barNdcbh = 1$, above which each ACH hosts more than one DCBH on average.}
\label{fig:M-Ndcbh_z10}
\end{figure}

\subsection{Occupation fraction at $z=10$}
\label{sec:barNdcbh_z10}

Here, we estimate the halo occupation fraction of DCBHs as a function of the halo mass from our model.
The occupation fraction is usually defined as the fraction of halos that host at least one DCBH.
In our model, we find that some halos host multiple DCBHs at different epochs in their progenitors as indicated by the thick line segments in Figs. \ref{fig:tZJ_SR0_data_4_0_0_609_7403490}
and \ref{fig:tZJ_SR1_data_4_0_0_609_7403490}.
Not to lose the information of the multiplicity, we here derive the average number $\barNdcbh$ of DCBHs per ACH.
Fig. \ref{fig:M-Ndcbh_z10} shows $\barNdcbh$ as a function of halo mass at the final redshift $\zfin = 10$. 
Here, in each logarithmic mass bin with an interval of 0.5 dex, we calculate the average numbers of ACHs and DCBHs by weighting the numbers in each zoom-in region by its corresponding overdensity probability (Eq. \ref{eq:od}). 
When $\Zth \leq 10^{-4}~\Zsun$, the number of DCBHs $\barNdcbh$ is smaller for {\SRi} (solid) than for {\SRo} (dashed)
as discussed in Section \ref{sec:effect_of_internal_radiation}.
$\barNdcbh$ increases with increasing $\Mhalo$ as
more massive halos tend to have a larger number of progenitor ACHs.
On average, one DCBH forms per ACH with a mass of $\Mhalo = 2 \times 10^{9}$ and
$6 \times 10^{9}~\Msun$ in the {\SRo} and {\SRi} models, respectively, depending slightly on the value of $\Zth$. 
When $\Zth = 10^{-3}~\Zsun$, 
$\barNdcbh$ becomes larger for {\SRi} than for {\SRo} at a given halo mass as also seen in Section \ref{sec:effect_of_internal_radiation}.
In this case, $\barNdcbh$ reaches unity at a halo mass of $\sim 10^{9}~\Msun$.
With this threshold halo mass, cosmological simulations showed that the formation of high-$z$ quasars can be explained \citep[e.g.,][]{Bhowmick22}.
This supports our estimate of minimum halo mass for DCBH formation.

For halos with masses $\gtrsim 10^9$--$10^{10}~\Msun$, $\barNdcbh$ exceeds unity, meaning that on average they can host multiple DCBHs.
However, those DCBHs may experience merger with each other along the merger tree when progenitors that host DCBHs merge with another progenitor or the main halo \citep{Chen22, Bhowmick22}.  
As we do not consider the merger of DCBHs in this study, the value shown in Fig. \ref{fig:M-Ndcbh_z10} should be taken as an upper limit.
This is further discussed in Section \ref{sec:multiBH}.


\subsection{Occupation fraction at the present time}
\label{sec:barNdcbh_z0}

It would be straightforward to compare our prediction with the observed halo occupation fraction of SMBHs in the local Universe \citep{Miller15} if we could follow the halo merger history down to $z=0$.
With the small side length of our zoom-in regions ($\sim 2~\Mpc$), however, halos that eventually form at $z=0$ would accrete matter also from outside the zoom-in regions.
For this reason, a simple extension of the simulations to $z=0$ is not of great value, and thus
we terminate our simulations at $\zfin = 10$. 

\begin{figure}
\includegraphics[width=\columnwidth]{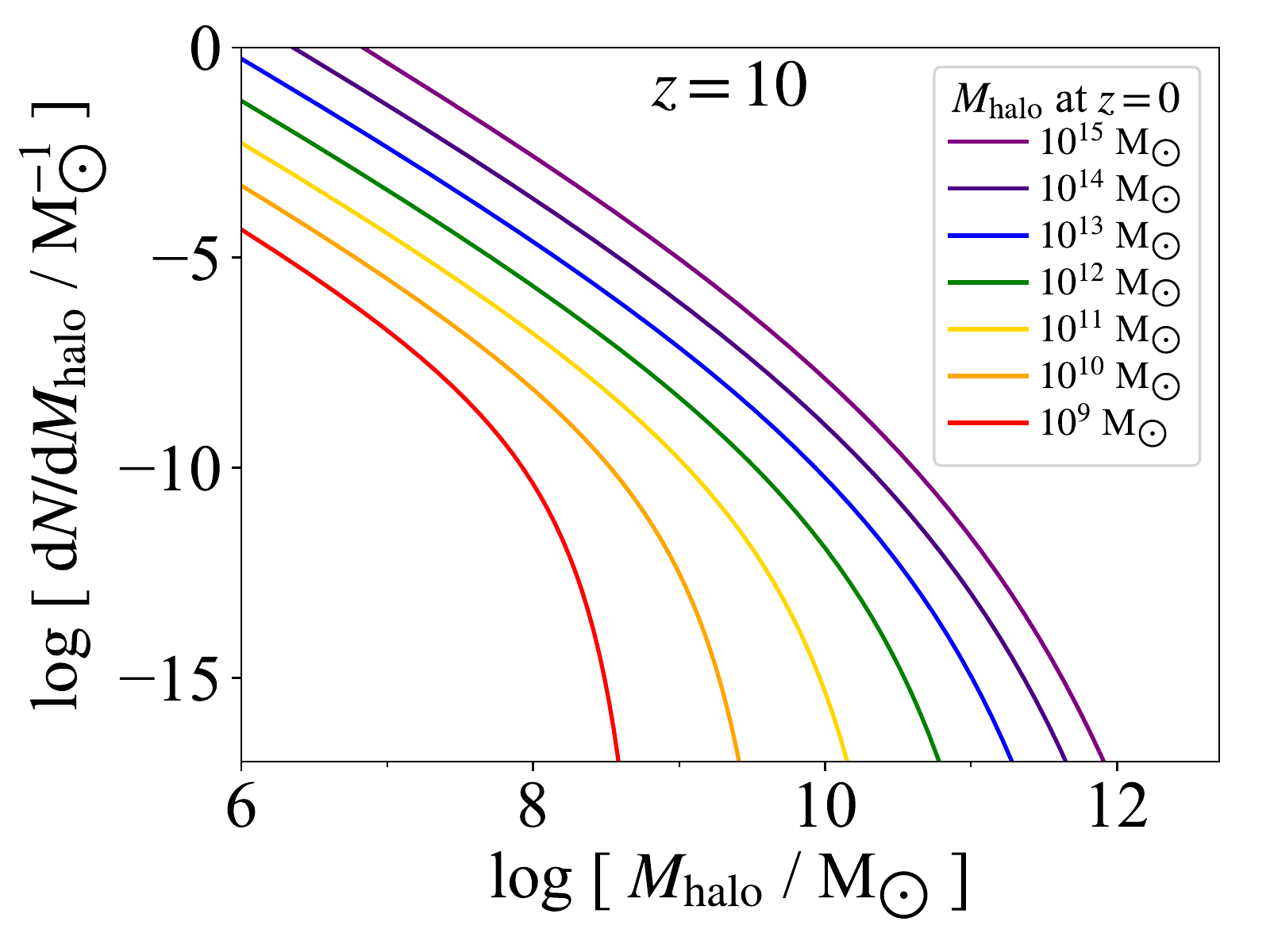}
\caption{
Mass distribution of progenitor halos at $z = 10$ derived from the EPS theory. The different colored curves correspond to progenitors of  final recipient halos at $z = 0$ with masses ranging from $10^9$ to $10^{15}~\Msun$ (from bottom to top; see the legend in the figure).}
\label{fig:MN_tab}
\end{figure}

\begin{figure}
\includegraphics[width=\columnwidth]{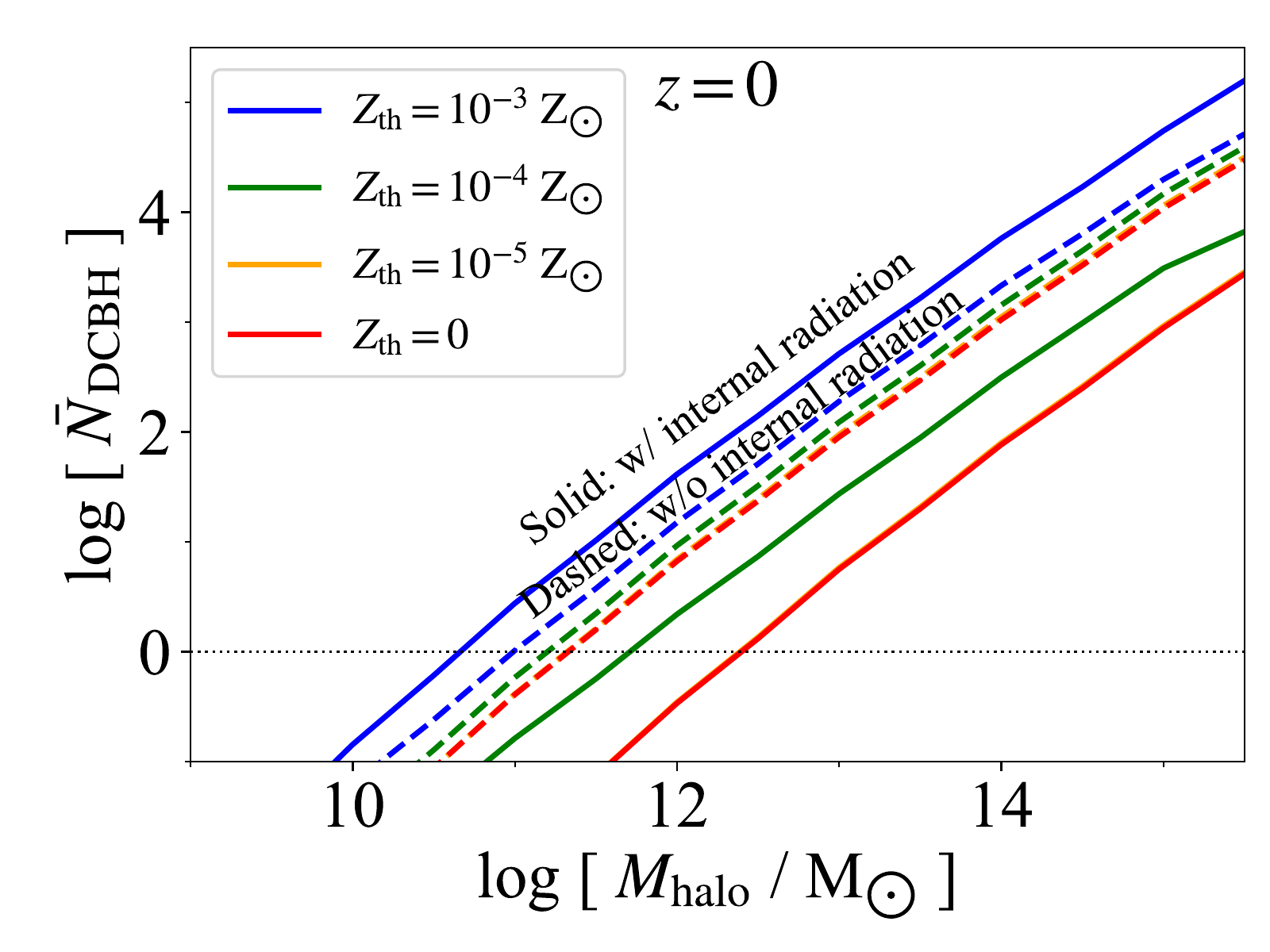}
\caption{The average number $\bar N_{\rm DCBH}$ of DCBHs hosted by a halo 
with mass $\Mhalo$ at redshift $z=0$.
The dashed and solid curves indicate the results without and
with internal radiation, respectively.
The red, orange, green and blue curves indicate the results
for threshold metallicities $\Zth = 0$, $10^{-5}$, $10^{-4}$ and $10^{-3}~\Zsun$, respectively.}
\label{fig:M-Ndcbh_z0}
\end{figure}

Instead, we adopt 
the following approach to estimate the average number $\barNdcbh$ of DCBHs hosted by a halo at redshift $z=0$.
For a halo with a given mass at $z=0$, we first obtain the mass distribution of its progenitors at $z=10$ by using the extended Press-Schechter (EPS) theory \citep{Bond1991} as shown in Fig. \ref{fig:MN_tab}.
We then integrate $\barNdcbh$ at $z=10$ (Fig. \ref{fig:M-Ndcbh_z10}) over the mass distribution.
Fig. \ref{fig:M-Ndcbh_z0} shows $\barNdcbh$ at $z=0$ as a function of
halo mass $\Mhalo$.
Although $\barNdcbh$ depends on whether the internal radiation is included or not and on the value of $\Zth$,
$\barNdcbh$ exceeds unity at masses $\Mhalo \gtrsim 10^{11}$--$10^{12}~\Msun$ in all the cases.
Assuming that the stellar mass is typically $\sim 0.01$ of the halo mass \citep{Moster10}, 
this indicates that the occupation fraction exceeds unity at stellar masses of $\Mstar \gtrsim 10^{9}$--$10^{10}~\Msun$, which
is consistent with the observations of SMBH occupation fraction in halos \citep[e.g.,][]{Miller15}. 



\section{Discussion}
\label{sec:discussion}

In this section, we first compare our results with earlier studies in Section \ref{sec:comparison}.
We then discuss the main limitations of our model in Section \ref{sec:caveats}.

\subsection{Comparison with other studies}
\label{sec:comparison}

Several authors have studied the origin of SMBHs, based on the DC scenario \citep{Dijkstra08, Agarwal12, Johnson13, Chon16, Valiante16, Habouzit16, Li21, Ni22, Trinca22, Spinoso22, Toyouchi22}.
These studies have also derived the number density $\nDCBH$ of DCBHs in a cosmological volume.
In this section, we compare the results of \citetalias{Chon16} and \citet[][hereafter \citetalias{Trinca22}]{Trinca22} with our work.

We first consider the results of \citetalias{Chon16}.
\citetalias{Chon16} estimated $\nDCBH = 0.003~\Mpc ^{-3}$ by performing $N$-body simulations
and semi-analytic calculations.
They focused on DCBH formation only in pristine halos in overdense regions ($3$--$4\sigma$) to explain the formation of high-redshift quasars.
Also, they did not consider internal radiation.
To check the consistency with our findings,
we estimate $\nDCBH$ only in the most overdense regions  ({\tt H1}--{\tt 4}) with $\Zth = 0$ in {\SRo}. 
Under these conditions, we obtain $\nDCBH = 0.01~\Mpc ^{-3}$, which is larger than \citetalias{Chon16}'s estimate by a factor of three.
This is mainly attributable to difference of halo finding algorithms.
\citetalias{Chon16} used {\sc subfind} \citep{Springel01} while we use {\sc rockstar} \citep{Behroozi13a}.
Although both of these methods can identify subhalos within the virial radius of the main halo, {\sc rockstar} can identify even subhalos spatially overlapping with the main halo, but with a finite relative velocity 
by using the information in the six-dimensional phase space.
As a result, the number of subhalos is larger than with {\sc subfind}.

\begin{figure}
\includegraphics[width=\columnwidth]{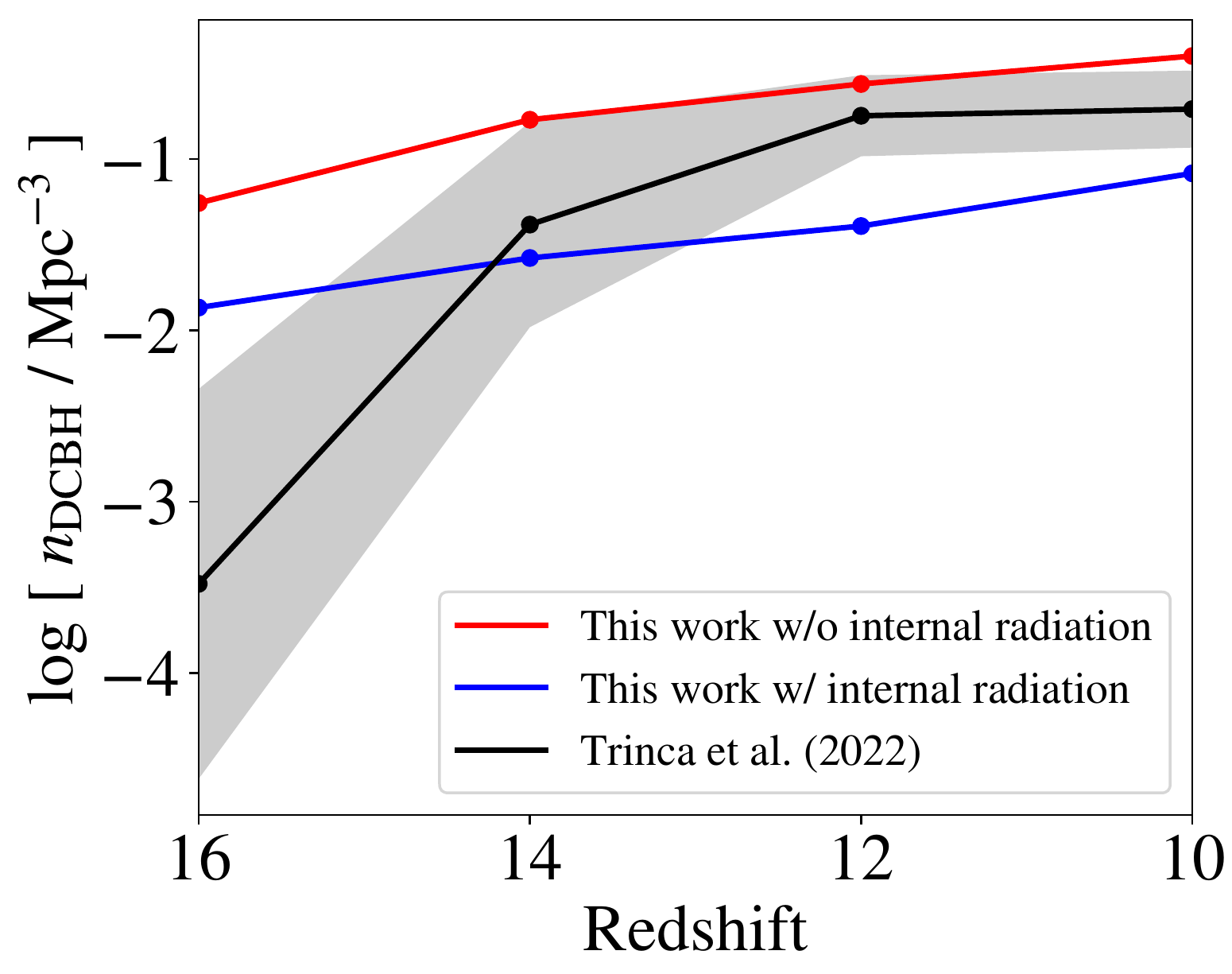}
\caption{The number density $\nDCBH$ of DCBHs as a function of the redshift.
The red and blue curves show the results in this work in cases without and with internal radiation, respectively, and the black curve shows the results of \citetalias{Trinca22}.
In this comparison, we impose the criteria for DCBH formation: (a) $\Tvir > 10^4$ K, (b) $\Jcr = 300$ and (c) $\Zth = 10^{-3.8}~\Zsun$,
the same as \citetalias{Trinca22}.
Note that the estimate of \citetalias{Trinca22} does not consider internal radiation.}
\label{fig:compa_T22}
\end{figure}

\citetalias{Trinca22} investigated the formation and growth of DCBHs with the semi-analytic calculation code {\sc cat} (Cosmic Archaeology Tool).
The black curve of Fig. \ref{fig:compa_T22} shows $\nDCBH$ as a function of redshift with a $1\sigma$ standard deviation (grey shaded area).
We recompute $\nDCBH$ with the same DC criterion as \citetalias{Trinca22}, i.e., $\Tvir > 10^4$ K, $\Zth = 10^{-3.8}~\Zsun$ and $\Jcr = 300$.
First, we confirm the consistency of their and our models with our {\SRo} run (the red curve), because \citetalias{Trinca22} did not include internal radiation.
At high redshifts $z\gtrsim 14$, $\nDCBH$ estimated by \citetalias{Trinca22} is smaller than our result (by two orders of magnitude at $z=16$). 
This is because \citetalias{Trinca22} assumed a uniform LW intensity produced by halos in each merger tree, and thus underestimate the number of DCBHs that form thanks to strong spatial fluctuations in the LW background intensity.
At $z\lesssim 14$, their results become consistent with our results within the errors. 
As more cosmic structures grow with time, fluctuations in the LW intensity decreases and the assumption of a uniform background is more realistic.
When we include internal radiation (the blue curve in Fig. \ref{fig:compa_T22}) with the same threshold metallicity as \citetalias{Trinca22}, $\Zth = 10^{-3.8}~\Zsun$, $\nDCBH$ becomes smaller by one order of magnitude for the same reason as discussed in Section \ref{sec:dependency_of_nDCBH}.
We then remark that neglecting the effect of internal radiation leads to overestimate the number density of DCBHs when $\Zth < 10^{-3}~\Zsun$.

\subsection{Limitations of our model}
\label{sec:caveats}

In this section, we discuss the limitations of our numerical model;
the absence of BH mass growth (Section \ref{sec:mass_growth}),
the possibility of tidal disruption of merging satellites (Section \ref{sec:tidal}) and 
dependence on model parameters (Section \ref{sec:model_parameters}).

\begin{figure}
\includegraphics[width=\columnwidth]{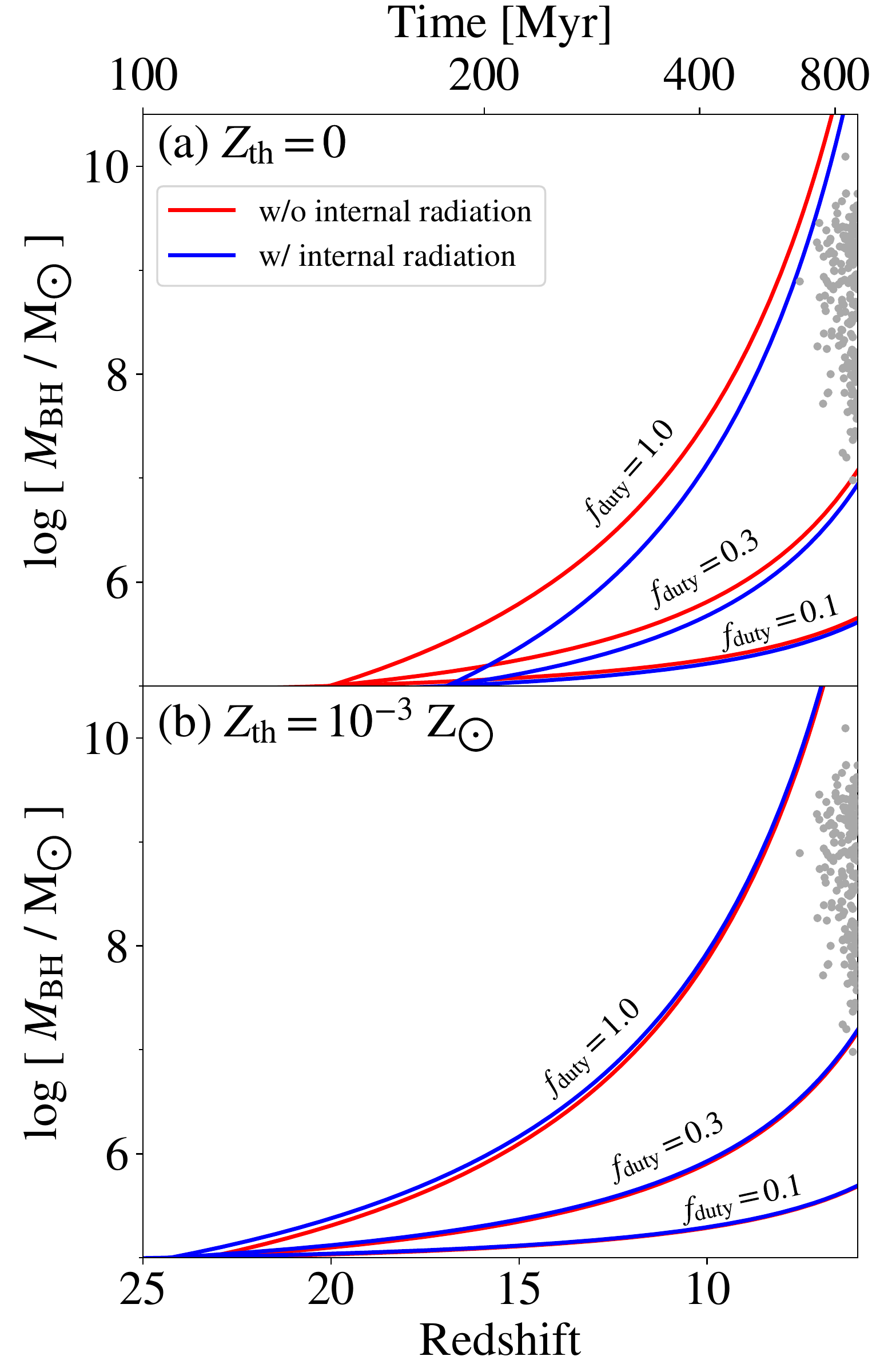}
\caption{The mass growth history of the DCBH that forms first in our runs without (red curves) and with (blue curves) internal radiation for (a) $\Zth = 0$ and (b) $10^{-3}~\Zsun$.
We assume that the DCBH has an initial mass of  $M_{\rm BH,0} = 10^5~\Msun$ at its formation redshift and grows at the Eddington rate.
With a fixed radiative efficiency $\epsilon = 0.1$, we plot the three cases with duty cycles of $\fduty = 0.1$, $0.3$ and $1$ from bottom to top. 
The grey dots represent the observational results of high-redshift quasars \citep[][and references therein]{Inayoshi20}.}
\label{fig:z-Mdcbh}
\end{figure}

\subsubsection{Mass growth of SMBHs}
\label{sec:mass_growth}

In our model, we do not consider the mass growth of DCBHs after their formation.
If BHs somehow find themselves in an environment with abundant gas reservoir, e.g., near the galactic center, they may grow rapidly in mass through gas accretion \citep{Valiante16, Li21, Ni22, Trinca22}, 
as it is required by models that aim to explain the formation of SMBHs at $z > 6$ (see \citealt{Inayoshi20} for a review).
In this work, we find that some DCBHs can form at very high redshifts $\sim 25$ when $\Zth = 10^{-3}~\Zsun$.
Assuming that they can grow in mass by accreting gas at the Eddington rate, these DCBHs can be the seeds of the largest SMBHs observed at $z > 6$. 

We plot in Fig. \ref{fig:z-Mdcbh} the mass growth of the first DCBHs that form in models with (a) $\Zth = 0$ (top panel) and (b) $10^{-3} Z_{\odot}$ (bottom panel) with and without internal radiation (blue and red lines, respectively), assuming Eddington-limited accretion.
In this rapid accretion regime, a BH can grow with an $e$-folding time
\begin{equation}
t _{\rm Edd} = \frac{450 \epsilon}{\fduty (1 - \epsilon)}~{\rm Myr} 
\end{equation}
with a radiative efficiency $\epsilon$ and duty cycle $\fduty$.
The two parameters, $\epsilon$ and $\fduty$, are poorly constrained and may depend on the BH accretion rate and redshift \citep{Haiman04, Shankar10, Shankar13, DeGraf17}.
For simplicity, we model the BH growth with a fixed $\epsilon = 0.1$ in a range of $\fduty = 0.1$--$1$ to estimate the parameter range where the DCBHs can reach the mass of observed high-redshift SMBHs (grey dots).

When DCBH formation is allowed only in the pristine gas ($\Zth = 0$), the first DCBH forms at $z=16$ thanks to internal radiation (blue curves in Fig. \ref{fig:z-Mdcbh}a),
which is later than in the mode without internal radiation ($z=21$).
This is because star formation is self-regulated by internal radiative feedback, and the stellar mass and LW emissivity becomes smaller (see Section \ref{sec:effect_of_internal_radiation}).
With $\fduty < 0.3$, even the BH that forms first cannot reach the mass of any observed SMBHs plotted in this Figure.
If it maintains Eddington-limited accretion with $\fduty \simeq 1$, the BH mass reaches
even the most massive SMBHs.
This indicates that $\fduty \simeq 1$ is required to explain the most massive BH formation if we allow DC to occur only in pristine halos.
The results are almost similar in the cases with $\Zth = 10^{-5}$ and $10^{-4}~\Zsun$.
With $\Zth = 10^{-3}~\Zsun$, the first BH forms around $z=25$ in both models with or without the internal radiation.
With duty cycle as low as $\fduty = 0.3$, 
the BH mass barely reaches $\sim 10^7~\Msun$ by $z=6$, and only the least massive among the observed SMBHs can be explained by our model.
With $\fduty = 1$, on the other hand, 
the first BH in our model can grow even more than the most massive one by this epoch.
Roughly speaking, $\fduty \ga 0.8$ is required for our first BHs to reach the observed most massive SMBHs at $z>6$.

We have explored the parameter range where observed high-redshift quasars can form in our DCBH model.
From the observational side, the most massive, high-redshift active SMBHs are rare objects ($\sim 1~{\rm Gpc}^{-3}$).
If all the DCBHs that form by $z=16$ grow to quasars, the number density will be $0.008$--$0.5$ ($0.05$--$0.07$) $~\Mpc ^{-3}$ in the case of {\SRi} 
({\SRo}, respectively).
This indicates that only a very small fraction ($10^{-9}$--$10^{-7}$) of DCBHs grow to bright highest-redshift quasars.
According to cosmological radiation hydrodynamics simulations, it is extremely hard that 
BH growth at the Eddington rate
is maintained all the time because of radiative feedback from the BH/nearby stars and of the large relative velocity between the BH and ambient gas reservoir \citep{Dubois15, Latif18, Chon21, Massonneau22, Sassano22}.
In addition, some simulation studies have demonstrated that BH dynamics can affect gas accretion in the early stage of BH mass growth \citep[e.g.,][]{Pfister19, Beckmann22}.
\citet{Pfister19} found that irregular distribution of stars in dwarf galaxies causes stochastic dynamical friction.
If the seed mass is on the order of $10^{4}~\Msun$, the orbit of the BHs are perturbed, and
BHs can migrate to low-density regions far from the galaxy center.
This would suppress mass accretion.
\citet{Beckmann22} found that BHs cannot grow sufficiently when their host galaxy is less massive $\lesssim 3\E{9}~\Msun$ due to the dynamical effect.

\subsubsection{The fate of multiple DCBHs in a halo}
\label{sec:multiBH}

Our result indicates that massive halos ($\Mhalo \gtrsim 10^{11}$--$10^{12}~\Msun$) can host multiple DCBHs (Fig. \ref{fig:M-Ndcbh_z0}).
This is because the DCBHs that have formed in satellite halos eventually gather in the main halo when the satellites merge with it.
One possible outcome is that those DCBHs merge with each other as they sink in the bottom of the halo potential well \citep{Chen22, Bhowmick22}. 
Another possibility is that the DCBHs formed in the satellites remain orbiting in the outskirt of the main halo without merger \citep{Regan23}.
The presence of off-centered active galactic nuclei (AGNs)
observed in present-day/high-redshift dwarf galaxies \citep[e.g.,][]{Webb12, Mezcua19} may be supporting the latter possibility.
\subsubsection{Tidal disruption of merging satellites}
\label{sec:tidal}

We have found that a fraction of halos meet the DC criterion thanks to the
external radiation from a more massive halo even in the {\SRi} runs.
Super-critical LW intensity is reached when the halos approach the main halo's virial radius (see Figs. \ref{fig:tZJ_SR0_data_4_0_0_609_7403490}
and \ref{fig:tZJ_SR1_data_4_0_0_609_7403490}).
Such close approach might cause their disruption by the tidal force of the main halo.
For example, hydrodynamical simulations performed by \citetalias{Chon16} show that 40 out of 42 halos selected as potential DC sites by
a semi-analytic model 
are found to be tidally disrupted before forming SMSs. 
We plan to examine whether the candidate DC halos identified in this study can host the formation of SMSs in spite of the tidal disruption by conducting hydrodynamics simulations.

\subsubsection{Dependence on model parameters}
\label{sec:model_parameters}

\noindent{(a) {\it Threshold time for high-UV duration}}
\vspace{0.1cm}

We have required halos to 
remain illuminated above $\Jcr$
longer than the threshold time $\tth = 4$ Myr, corresponding to the
free-fall time of a cloud with a density of $100~\percc$, to induce the DCBH formation.
If we adopt a longer threshold time of $\tth > 4$ Myr, i.e., more strict condition for DCBH formation, the number of DCBHs will be smaller.
Since the typical lifetime of massive stars is $\sim 4$ Myr, metal enrichment from SNe will take place in the meantime.
This may boost the metallicity in ACHs up to $> 10^{-3}~\Zsun$ and prohibit DCBH formation.
The actual UV irradiation duration required for the DC would depend on the density of a DCBH forming clump in the halo.
From the $N$-body simulations alone, we cannot derive the gas density of clumps in halos.
We plan to study the fate of those DC halos by hydrodynamics simulations in a forthcoming paper.

\vspace{0.3cm}
\noindent{(b) {\it Threshold LW intensity}}
\vspace{0.1cm}

The critical LW intensity $\Jcr$ depends on the SED of radiation sources 
\citep{Omukai01}, as well as on the detailed rate coefficients of relevant chemical reactions \citep{Shang10}.
We have used the critical LW intensity of $\Jcr = 10^3$ taken from
\citet{Sugimura14}, who calculated the $\Jcr$ for realistic SEDs
of star-forming galaxies with metallicities $0$--$1~\Zsun$ 
\citep{Leitherer99, Schaerer03}.
Previously,  \citetalias{Chon16} applied $\Jcr = 10^4$ and $10^2$ for
Pop III and II sources, respectively, referring to the result of three-dimensional simulations of \citet{Shang10},
who used an updated H$_2$ collisional dissociation rate \citep{Martin96}.
If we adopt a lower value $\Jcr = 10^2$ for Pop II sources as in \citetalias{Chon16} than our fiducial value ($\Jcr = 10^3$), more halos will satisfy the DC criterion, and the average number of DCBHs $\barNdcbh$ will be higher by an order of magnitude both at redshifts $z=10$ and $z=0$.

\section{Conclusion}
\label{sec:conclusion}

We have studied the number density of direct-collapse black-holes forming in the early universe by means of 
cosmological $N$-body simulations coupled with a semi-analytic model of galaxy evolution. 
We find that halos with masses $\gtrsim 10^{9}$--$10^{10}~\Msun$
host more than one DCBH on average at the final redshift $z=10$, the last snapshot of our simulations.
After hierarchical mergers of the host halos, those DCBHs are incorporated into more massive halos.
We estimate that halos with masses $\gtrsim 10^{11}$--$10^{12}~\Msun$ 
host more than one DCBH at the present epoch by using the extended Press-Schechter model.
Since the stellar-to-halo mass ratio is typically $\sim 0.01$ in the halo mass range \citep{Moster10},
our estimate is consistent with the observational halo occupation
fraction in the local Universe that approaches unity above a stellar masse of $\sim 10^9~\Msun$ \citep[e.g.,][]{Miller15}.
This indicates that the DC scenario alone may be able to explain the origin of the majority 
of observed SMBHs.

\section*{ACKNOWLEDGMENTS}

We thank T. Hosokawa, K. Inayoshi and D. Toyouchi and C. Kobayashi for fruitful discussions.
The numerical simulations and analyses in this work are carried out on
XC40 in Yukawa Institute of Theoretical Physics (Kyoto University) through the courtesy of Prof. Kunihito Ioka, and
XC50 {\sc aterui II} in Oshu, Iwate at the Center for Computational Astrophysics (National Astronomical Observatory of Japan) through the courtesy of Prof. Eiichiro Kokubo.
This research is supported by Grants-in-Aid for Scientific Research (KO: 17H06360, 17H01102, 17H02869, 22H00149) from the Japan Society for the Promotion of Science.  
We did most of the analysis with {\sc yt} \citep{yt}.
The figures in this paper are constructed with the
plotting library {\sc matplotlib} \citep{matplotlib}.
KO acknowledges support from the Amaldi Research Center funded by the MIUR program "Dipartimento di Eccellenza" (CUP:B81I18001170001).

\section*{Data availability}

The simulation data will be shared on reasonable request to the authors.


%
%


\label{lastpage}

\end{document}